# Conformation-dependent sequence design of polymer chains in melts

Elena N Govorun,* Ruslan M Shupanov, Sophia A Pavlenko, Alexei R Khokhlov

Faculty of Physics, M. V. Lomonosov Moscow State University, Leninskie gory, Moscow 119991 Russia
E-mail: govorun@polly.phys.msu.ru

**Abstract**
Conformation-dependent design of polymer sequences can be considered as a tool to control macromolecular self-assembly. We consider the monomer unit sequences created via the modification of polymers in a homogeneous melt in accordance with the spatial positions of the monomer units. The geometrical patterns of lamellae, hexagonally packed cylinders, and balls arranged in a body-centered cubic lattice are considered as typical microphase-separated morphologies of block copolymers. Random trajectories of polymer chains are described by the diffusion-type equations and, in parallel, simulated in the computer modeling. The probability distributions of block length $k$, which are analogous to the first-passage probabilities, are calculated analytically and determined from the computer simulations. In any domain, the probability distribution can be described by the asymptote ~ $k^{-3/2}$ at moderate values of $k$ if the spatial size of the block is less than the smallest characteristic size of the domain. For large blocks, the exponential asymptote $\exp(-\,\mathrm{const}\; ka^2 / d_{\mathrm{as}}^2)$ is valid, $d_{\mathrm{as}}$ being the asymptotic domain length ($a$ is the monomer unit size). The number average block lengths and their dispersities change linearly with the block length for lamellae, cylinders, and balls, when the domain is characterized by a single characteristic size. If the domain is described by more than one size, the number average block length can grow nonlinearly with the domain sizes and the length $d_{\mathrm{as}}$ can depend on all of them.

## 1. Introduction

Polymer sequences can control the conformation of single macromolecules, as well as the spatial structure of polymer solutions and melts. This ability is obviously revealed for biomacromolecules with the unique native structure of proteins and the specific conformations of DNA and RNA [1-4]. The attempts to design artificial biomimetic polymers have led, in particular, to the idea of the conformation-dependent design of protein-like copolymers with hydrophobic core and polar shell. In a homopolymer globule consisting of hydrophobic monomer units, the monomer units in the surface layer are transformed into polar ones. The macromolecules thus obtained possess the ability to form soluble globules, which are stable against aggregation with each other.

First protein-like copolymers were designed and described in the computer experiment using the Monte Carlo method [5]. Then, the statistical properties of sequences and the globule stability were investigated in the theory and computer simulations [6-8], and the higher stability of polymer globules formed after the conformation-dependent chemical modification was demonstrated experimentally [9, 10]. Besides, the evolutionary conformation-dependent process of mutations (monomer unit exchange) was studied in the theory and computer simulations, the final macromolecular conformations were hydrophobic-polar core-shell structures and core-tail (tadpole) conformations [11].

In addition to the sequences of core-shell globules, the conformation-dependent sequence design was theoretically described for the macromolecules adsorbed at a penetrable liquid-liquid interface [12]. For homopolymers adsorbed at a solid plane surface, the conformation-dependent sequence modification was carried out in the computer simulations, and the adsorption of macromolecules with the different types of monomer unit distributions was studied both in silico and experimentally [13, 14].

Aside from single macromolecules, all chains of a melted polymer can be modified. It is microphase separation that plays an important role in the material properties of block copolymers. The simplest lamellar structure was used as a pattern for the conformation-dependent sequence design in the homopolymer melt, and the statistical characteristics of sequences were studied in the theory and computer simulations. Using the dissipative particle dynamics, the lamellar-type structure was reproduced in the melt of the designed copolymers after switching on the repulsive interactions between the monomer units of different types [15].

The macromolecules obtained via the conformation-dependent modification can be considered as random multiblock copolymers, where a block is a sequence of monomer units of the same type bounded by the monomer units of another type. The monomer unit sequences can be characterized by the average composition and the block length distributions. At present, a great progress in the synthetic methods for multiblock copolymers with controlled length and degree of blockiness attracts attention to these materials [16-20].

In a large homogeneous globule and in a melt, the polymer chain conformations are random, and the dependence of the monomer coordinates on its number along the chain can be reduced to the trajectory of a random walker. The statistical weight of chain trajectories can be described by the Green function obeying the diffusion-type equation in terms of the Gaussian chain model [21]. For hydrophobic blocks in the globule core of a protein-like copolymer, the block termination at the core surface is determined by the absorbing boundary conditions for the diffusion equation [6]. For copolymer blocks corresponding to the lamellar pattern in melts, the block length distribution is also determined by the absorbing boundary conditions.

In the probability theory, such analysis relates to the classical first-passage problem [22, 23]. Generalization of the problem to the cases of the moving walls or the random walks of special types continues to attract the interest [24-29]. In the present paper, the block length distribution and the number average block length correspond to the first-passage probability and the first-passage time in the 3D domains of different shapes.

For the polymer blocks in a globule core or in a lamella [6, 15], the asymptote for the distribution of block length, $k$, is $\sim k^{-3/2}$ at moderate values of $k$, and such a power-law distribution can be described in terms of the Levy-flight statistics [30]. At very large values of $k$, the probability distribution of block lengths decreases exponentially. For protein-like copolymers, the wide block length distributions lead to the existence of long-range correlations in the sequences [6].

In the present paper, we put together the descriptions of sequence statistics for the alternating lamellae and periodically arranged cylinders and balls. The most detailed analysis is performed for the first considered pattern of hexagonally packed cylinders. In the section Model and methods, the system, the patterns, and the model approximations are described, the diffusion equation with the boundary conditions and the asymptotes are written out, and the dissipative particle dynamics (DPD) model is presented. In the section Results and Discussion, the exact solutions of the diffusion equation in the considered domains are written out, the smoothing procedure for the computer simulation data is explained, the characteristics of the block length distributions are calculated, and their peculiarities are analyzed. In the Conclusions section, the results of the work are summarized.

## 2. Model and methods

Initially, we consider a homogeneous melt of homopolymer molecules each consisting of $N$ monomer units. The polymer chains are assumed to be ideal and their conformations are described as random walks with the mean-square distance $a$ between two neighbor monomer units in a chain.

In that case, the distribution of distances between any two monomer units of the same polymer chain is described by the Gaussian statistics [21].

Then, we introduce an imaginary two-component bulk pattern that resembles one of the typical morphologies of microphase separation in block copolymer melts. Those morphologies are alternating lamellae, hexagonally arranged cylinders, and balls forming a cubic lattice, in particular, a body-centered cubic lattice (figure 1). Type A is assigned to the monomer units in the yellow (light) lamellae, cylinders and balls, whereas type B is assigned to all other monomer units in the blue (dark) domains. The sequence of morphologies in figure 1 follows a decrease in the volume fraction of yellow domains and, correspondingly, decreasing in the mean fraction of monomer A units in a polymer chain. When moving along a polymer chain, the type of monomer units changes at the interfaces between blue and yellow domains. Thus designed chains can be described as random multiblock copolymers, the chains of which consist of blocks, i.e., sequences of chemically identical monomer units.

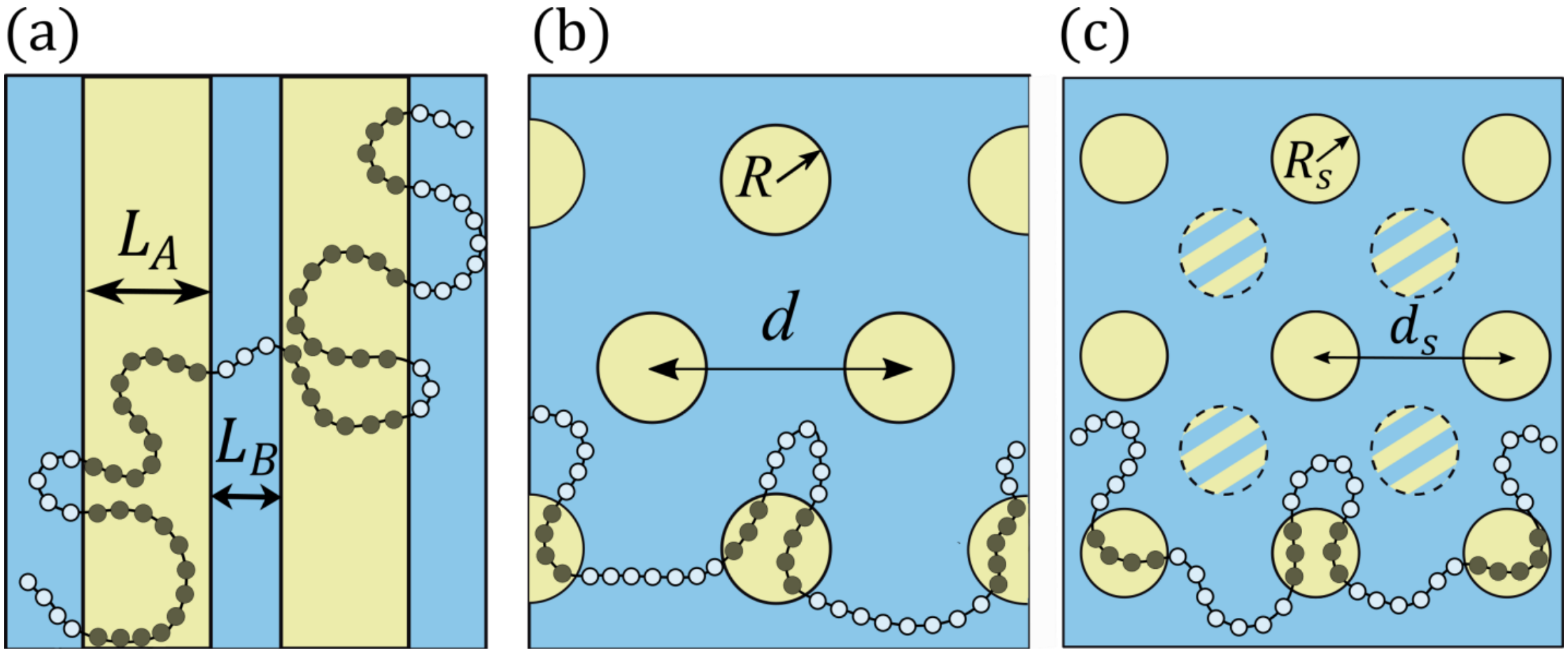

Figure 1. Scheme of modification of an initially homopolymer melt into a patterned melt of multiblock copolymer. Monomer units in the yellow (light) lamellae, in the cylinders and spheres are of type A, other monomer units are of type B. The cross-sections of lamellae (a), hexagonally packed cylinders (b), and a body-centered cubic lattice of balls (c) are shown. For the last pattern (c), the projections of the inner balls of elementary cells onto the section plane are shown by hatching.

Monomer unit distribution along a polymer chain can be completely described by the distributions of A and B blocks over their length together with the average chain composition. In the work, those statistical characteristics of monomer unit sequences are determined via two independent methods. The first one is based on the calculation of the Green function of a polymer chain in the considered domains via solving the diffusion-type equations. The second one uses computer simulations of a homogeneous polymer melt with the geometrical patterns (figure 1), where the number of blocks of each length is directly counted.

*2.1 Sequence statistics via random walk analysis*

The use of the diffusion-type equations for the description of polymer conformational statistics is based on the analogy between a random walk trajectory and a dependence of the position of a monomer unit on its curvilinear coordinate along an ideal chain. The statistical weight of chain conformations with fixed ends is described by the Green function of a polymer chain [21].

Polymer block trajectories begin at interfaces between A (yellow) and B (blue) domains and finish somewhere at those interfaces. Further, a domain (or monomer unit) type is denoted by $\alpha$; $\alpha$ = A,B. The Green function $G_\alpha(\mathbf{r},k|\mathbf{r}_0)$ of the sequences containing $k$ monomer units of type $\alpha$, beginning at the point $\mathbf{r_0}$ and ending at the point $\mathbf{r}$ in the $\alpha$ domain obeys the equation

$$\frac{\partial}{\partial k}G_\alpha(\mathbf{r},k\,|\,\mathbf{r}_0)=\frac{a^2}{6}\Delta G_\alpha(\mathbf{r},k\,|\,\mathbf{r}_0),\qquad \mathbf{r},\mathbf{r}_0\in V_\alpha, \tag{1}$$

where $\Delta$ is the Laplacian with respect to $\mathbf{r}$. The absorbing boundary condition describes the sequence termination if its trajectory hits the surface $S_\alpha$ of α domain:

$$G_\alpha(\mathbf{r},k\,|\,\mathbf{r}_0)\big|_{\mathbf{r}\in S_\alpha}=0\,. \tag{2}$$

The initial condition is

$$G_\alpha(\mathbf{r},0\,|\,\mathbf{r}_0)=\delta(\mathbf{r}-\mathbf{r}_0) \tag{3}$$

A part of the polymer chain in the α domain with both ends at the interfaces becomes an α block, and then the initial point $\mathbf{r}_0$ should be taken near the interface.

Note that macromolecules are assumed to be long enough, so that the probability of chain termination inside a domain can be neglected. This condition is hold for $Na^2 >> d_\alpha^2$, α = A,B, where $d_\alpha$ is the characteristic size of α domain.

For sequences of length $k$ with the characteristic spatial size $\sqrt{k}a$ being much smaller than a characteristic domain size $d_\alpha$, the Green function can be approximated by the solution of Eq. (1) in a semi-space $x \geq 0$:

$$G_\alpha(\mathbf{r},k\,|\,\mathbf{r}_0)=G_\perp(x,k\,|\,x_0)G_0(y,k\,|\,y_0)G_0(z,k\,|\,z_0), \tag{4}$$

where $G_0(y,k\,|\,y_0)=\sqrt{\dfrac{3}{2\pi ka^2}}\exp\left(-\dfrac{3}{2ka^2}(y-y_0)^2\right)$ is a univariate Gaussian distribution and $G_\perp(x,k\,|\,x_0)=G_0(x,k\,|\,x_0)-G_0(x,k\,|-x_0)$. In this approximation, the Green function (4) does not depend on the segment type $\alpha$. The characteristic domain size $d_\alpha = L_\alpha$ for lamellae, $d_A = R$ for cylinders, and $d_A = R_s$ and for balls (figure1).

We are aimed to determine the probability distributions $P_A(k)$ and $P_B(k)$ of A and B blocks, respectively, over their length $k$. The product $P_\alpha(k)\cdot\Delta k$ should give the fraction of blocks with the length in the interval [$k$, $k+\Delta k$]. It can be related to the integral $\int_{V_\alpha} d\mathbf{r}\, G_\alpha(\mathbf{r},k\,|\,\mathbf{r}_0)$, which is equal to the number of chain trajectories consisting of $k$ steps, beginning at the point $\mathbf{r_0}$ and staying inside the domain, in proportion to the total number of possible trajectories. Then, the probability distribution $P_\alpha(k)$ is equal to

$$P_\alpha(k) = -\frac{\partial}{\partial k}\int_{V_\alpha} d\mathbf{r}\, G_\alpha(\mathbf{r},k\,|\,\mathbf{r}_0) \tag{5}$$

or, with the use of Eq. (1), it can be calculated as an integral probability flux through the surface:

$$P_\alpha(k) = -\frac{a^2}{6}\oint_{S_\alpha} ds(\mathbf{n}\nabla G_\alpha), \tag{6}$$

where the unit vector $\mathbf{n}$ is perpendicular to the surface.

The absorbing boundary condition leads to $G_\alpha \to 0$ at $k \to \infty$. Then, from the formula (5) and the initial condition (3), the block size distribution $P_\alpha(k)$ should obey the normalization condition:

$$\int_0^\infty dk\, P_A(k) = 1 \tag{7}$$

For the solution (4) in a semi-space, the distribution of blocks over their length can be found with the use of Eq. (6):

$$P_\alpha(k) = \frac{a^2}{6}\frac{\partial G_\perp(x,k\,|\,x_0)}{\partial x}\bigg|_{x=0} = \sqrt{\frac{3}{2\pi k a^2}}\,\frac{x_0}{k}\exp\left(-\frac{3x_0^2}{2ka^2}\right) \tag{8}$$

This dependence can be interpreted also as a survival function describing the time of first passage of the plane $x = 0$ by the random walk [22].

A sequence of blocks in a polymer chain can be considered as a unilateral walk with the step size distribution described by the alternating functions $P_A(k)$ and $P_B(k)$. For the value $x_0$ being of the order of $a$ and for $k >> 1$, the dependence of $P_\alpha(k)$ on $k$ (Eq. (8)) can be approximated as

$$P_\alpha(k) \sim 1/k^{3/2}. \qquad 1 \ll k \ll (d_\alpha/a)^2 \tag{9}$$

A random walk with such a power-low distribution of step size (Eq. (9)) is often referred to as a Levy flight [30].

For larger blocks of the spatial size that is comparable with the characteristic size of the domain $d_\alpha$, this size and the shape of the interface become substantial. For very long α blocks with trajectories avoiding the boundaries, both the Green function and the block length distribution decrease exponentially with $k$. Having in mind that the distribution $P_\alpha(k)$ should be continuously extended over $k > k^* = (d_\alpha/a)^2$, the asymptotic form

$$P_\alpha(k) \sim \left(\frac{a}{d_\alpha}\right)^3 \exp\left(-\lambda_\alpha \frac{ka^2}{d_\alpha^2}\right), \quad k \gg (d_\alpha/a)^2 \tag{10}$$

can be written out, where the coefficient $\lambda_\alpha$ depends on the geometrical parameters of the region, the value of $\lambda_\alpha$ being of the order of unity. It is worth noting that the asymptotic form (10) describes the distribution $P_\alpha(k)$ for the case of a single characteristic domain size $d_\alpha$. If the shape of the domain is complex and it is characterized by several spatial sizes, the behavior of the function $P_\alpha(k)$ at large $k$ could depend on several parameters. The power-low dependence (9) at small $k$ is valid until the block spatial size $\sqrt{k}a$ becomes comparable with the smallest characteristic size of the domain.

Useful characteristics of a block length distribution are the number average length $\bar{k}_{n\alpha}$, mass average length $\bar{k}_{w\alpha}$, and dispersity $Đ_\alpha$:

$$\bar{k}_{n\alpha} = \int_0^\infty dk\, kP_\alpha(k), \quad \bar{k}_{w\alpha} = \frac{1}{\bar{k}_{n\alpha}} \int_0^\infty dk\, k^2 P_\alpha(k), \quad Đ_\alpha = \frac{\bar{k}_{w\alpha}}{\bar{k}_{n\alpha}} \tag{11}$$

The number average length is analogous to the first-passage time of a random walker. For the asymptotic forms (9) and (10), the dependences (11) of the average lengths on the characteristic size $d_\alpha$ of the domain are

$$\bar{k}_{n\alpha} \sim (d_\alpha/a), \quad \bar{k}_{w\alpha} \sim (d_\alpha/a)^2. \tag{12}$$

Then, the dispersity of the block length distributions

$$Đ_\alpha \sim (d_\alpha/a) \tag{13}$$

grows linearly with the domain size.

The exact solutions of Eq. (1) are known for a ball, a cylinder, and a lamella [30], whereas such solutions are unknown for an infinite domain with the excepted domains in the shape of cylinders or spheres (as the blue domains in figure 1 (b) and (c)). In diblock copolymer melts, the sizes of the domains of different types are usually comparable. In this case, the probability of trajectories that never hit absorbing boundaries is negligible. In the theoretical analysis, we approximate these infinite domains by the simplest finite ones. For B blocks beginning near a certain cylinder (the pattern (b) in figure 1), we approximate the domain by a cylindrical layer surrounding this cylinder. For B blocks with the beginning near a certain ball (the pattern (c)), the approximation of a spherical layer is used.

For some of the domain shapes in figure 1, the block length distributions for the pattern-modified polymers were calculated previously. Blocks in a ball can be described by the distribution function derived for hydrophobic blocks in a spherical core of a large globule formed by the protein-like copolymer [6]. Besides, the distribution of the polar block lengths in a protein-like copolymer was calculated [6], where an additional reflecting boundary condition at the outer surface was used for the spherical layer around the core, not as in the present consideration for B blocks. The block length distribution for a lamellar structure was written out and modeled in the computer simulations in the work [15].

*2.2 Computer simulations*

In computer simulations, we generate a homogeneous polymer melt, assign types A and B to the monomer units according to the geometrical patterns in figure 1, and find the block length distributions. Polymer chains are modeled as identical phantom beads (monomer units) connected by "springs". Initially, polymer chains are randomly distributed in the simulation box with periodic boundary conditions and the conformations of polymer chains are set as random walks with a constant spatial step. Then, the system of beads is equilibrated using the method of dissipative particle dynamics [32] until the conformational characteristics of polymer chains stop changing.

The coordinates of beads are described by the radius-vectors $\{\mathbf{r}_i\}$. The force $\mathbf{f}_i$ acting on the $i$-th bead in the system is a sum of the bond stretching forces $\mathbf{f}_{ij}^{\mathrm{b}}$ acting from the neighbor beads in the chain, the conservative forces $\mathbf{f}_{ij}^{\mathrm{c}}$, dissipative forces $\mathbf{f}_{ij}^{\mathrm{d}}$, and random forces $\mathbf{f}_{ij}^{\mathrm{r}}$:

$$\mathbf{f}_i = \sum_{j \neq i} \left( \mathbf{f}_{ij}^{\mathrm{b}} + \mathbf{f}_{ij}^{\mathrm{c}} + \mathbf{f}_{ij}^{\mathrm{d}} + \mathbf{f}_{ij}^{\mathrm{r}} \right) \tag{14}$$

In a Gaussian chain, the bond stretching force depends linearly on the distance between beads $i$ and $j$:

$$\mathbf{f}_{ij}^{\mathrm{b}} = -K\mathbf{r}_{ij}, \quad \mathbf{r}_{ij} = \mathbf{r}_i - \mathbf{r}_j \tag{15}$$

where the constant $K$ characterizes the bond stiffness. The forces $\mathbf{f}_{ij}^{\mathrm{c}}$, $\mathbf{f}_{ij}^{\mathrm{d}}$, and $\mathbf{f}_{ij}^{\mathrm{r}}$ are acting from all $j$-th beads in the sphere of the cut-off radius $r_c$ with the $i$-bead in the center. The conservative forces $\mathbf{f}_{ij}^{\mathrm{c}}$ are taken in the form

$$\mathbf{f}_{ij}^{\mathrm{c}} = \begin{cases} -\upsilon(1 - r_{ij})\mathbf{r}_{ij}/r_{ij}, & r_{ij} \le 1 \\ 0, & r_{ij} > 1 \end{cases} \tag{16}$$

where $\upsilon$ is the interaction parameter.

The bead mass $m$, the cut-off radius $r_c$, and the thermal energy $k_B T$ are taken as reference units. Expressed in those units, the stiffness coefficient $K = 4$, the initial bond length is equal to 1, the interaction coefficient $\upsilon = 25$, and the mean number density of monomer units is equal to 3. The dissipative forces $\mathbf{f}_{ij}^{\mathrm{d}}$ and random forces $\mathbf{f}_{ij}^{\mathrm{r}}$ are set as in the work [33] to provide the value of the noise parameter $\sigma = 3$. The modified velocity-Verlet algorithm is used for the integration over time with the time step of 0.04 [34].

For spherical particles packed into a body-centered-cubic lattice and for lamellae, the box size 60x60x60 is taken. For hexagonally packed cylinders, the box size is 60x $60\sqrt{3}$ x60. The polymer chains consist of 1000 beads each, the mean square distance between neighboring beads in a chain after equilibration is approximately equal to 1.03. Since the values of $a$ and cut-off distance $r_c$ are close to each other, below we use in the notation for the computer simulations the value $a$ as a unit length as if $a = r_c$. All considered simulation boxes contain an integer number of structure periods to satisfy the periodic boundary conditions.

After the equilibration, the types A and B are assigned to monomer units in accordance with their spatial positions. The monomer units in the yellow domains (lamellae, cylinders, and spheres in figure 1) are of type A and the others in the blue domains are of type B. Then, the numbers of blocks are found directly for all possible block lengths. The number and mass average block lengths are calculated and the block length distributions are plotted in comparison with the theoretical predictions.

## 3. Results and discussion

*3.1 Lamellar structure*

The blocks of type A and blocks of type B form alternating flat layers of thickness $L_A$ and $L_B$, respectively (figure 1a). Statistical characteristics of the block distribution of $\alpha$ type, $\alpha$ =A,B, are determined by the layer thickness $L_\alpha$ and the monomer unit size $a$. The solution of the equation (1) in a layer $0 \le x \le L_\alpha$ can be represented as a product $G_\alpha(\mathbf{r},k\,|\,\mathbf{r}_0) = G_{\alpha 1}(x,k\,|\,x_0)G_0(z,k\,|\,z_0)G_0(y,k\,|\,y_0)$, where the unidimensional Green function $G_{\alpha 1}(x,k\,|\,x_0)$ obeys the equation

$$\frac{\partial G_{\alpha 1}}{\partial k} = \frac{a^2}{6}\frac{\partial^2 G_{\alpha 1}}{\partial x^2}. \tag{17}$$

Its solution is

$$G_{\alpha 1}(x,k\,|\,x_0) = \frac{2}{L_\alpha}\sum_{j=1}\exp\left(-\frac{j^2\pi^2 a^2}{6L_\alpha^2}k\right)\sin\frac{j\pi x_0}{L_\alpha}\sin\frac{j\pi x}{L_\alpha} \tag{18}$$

Using equations (6) and (18), the block length distribution can be found:

$$P_\alpha(k) = \frac{2\pi}{3\tilde{L}_\alpha^2} \sum_{j=1,3,5,\ldots} j \exp\left(-\frac{j^2\pi^2}{6\tilde{L}_\alpha^2} k\right) \sin\frac{j\pi\tilde{x}_0}{\tilde{L}_\alpha}, \tag{19}$$

where $\tilde{L}_\alpha = L_\alpha / a$, $\tilde{x}_0 = x_0 / a$.

In computer simulations, all macromolecules were scanned from one end to another and the block lengths and their numbers $n_\alpha(k)$ were found in the system after the equilibration. The total number of blocks of type $\alpha$ is equal to $n_{\alpha(\mathrm{tot})} = \sum_k n_\alpha(k)$ and the block length distribution can be calculated as $\tilde{P}_\alpha(k) = n_\alpha(k) / n_{\alpha(\mathrm{tot})}$. If $k_{\alpha\,\max}$ is the length of the longest α block in the system, then $\tilde{P}_\alpha(k) = 0$ at $k > k_{\alpha\,\max}$.

If $k$ exceeds several tens the dependence $\tilde{P}_\alpha(k)$ fluctuates remarkably, and for large $k$ some block lengths are not found in the system. To find a smooth dependence at large values of $k$, we calculate the averaged fraction of blocks in an interval of length $\Delta k$ as

$$\tilde{P}_{\alpha(\mathrm{av})}(k) = \frac{1}{\Delta k} \sum_{i=k_0}^{k_0+\Delta k-1} \tilde{P}_\alpha(i), \tag{20}$$

where $k_0 = k - [\Delta k / 2]$.

The block length distribution $P_\alpha(k)$ (Eq. (19)) obtained via solving the diffusion-type equation is continuous and it changes sharply for quite small $k$ ($k < 3$). For comparing this dependence with the discrete one obtained in computer simulations, we transform the expression (19) into a discrete form $\tilde{P}_\alpha(k)$ as

$$\tilde{P}_\alpha(k) = \int_{k-1}^{k} dk' P_\alpha(k'), \tag{21}$$

for which the normalization condition is fulfilled for the sum:

$$\sum_{k=1}^{\infty} \tilde{P}_\alpha(k) = \int_0^\infty dk\, P_\alpha(k) = 1. \tag{22}$$

The block length distribution in a discrete form is

$$\tilde{P}_\alpha(k) = \frac{4}{\pi} \sum_{j=1,3,\ldots} \frac{1}{j} \sin\frac{j\pi\tilde{x}_0}{\tilde{L}_\alpha} \left( e^{-\frac{j^2\pi^2}{6\tilde{L}_\alpha^2}(k-1)} - e^{-\frac{j^2\pi^2}{6\tilde{L}_\alpha^2}k} \right), \quad k = 1,2,\ldots \tag{23}$$

The probability distributions $\tilde{P}_\alpha(k)$ hereafter are calculated numerically using the original program code in Free Pascal. The difference in the values $\tilde{P}_\alpha(k)$ and $P_\alpha(k)$ given by the expressions (19) and (21) is distinguishable only at $k < 5$. The values of $P_\alpha(k)$ are calculated for the block length from $k = 1$ to $k_{\alpha\,\max}^{(theor)}$ corresponding to the probability $P_\alpha(k)$ values $10^{-7}$–$10^{-6}$.

The accuracy of calculations is controlled by the value of the sum in the normalization condition (22), which is demanded to deviate from unity not larger than in 1-2%. The required number of terms in the sum (23) is less than ten at $k > 10$, whereas one hundred or more terms are necessary at $k = 1$.

In figure 2 the block length distributions for lamellae of thickness $L = L_A = 5a$ and $15a$ are presented in the double logarithmic coordinates. The simulated values $\tilde{P}_A(k)$ are shown by markers at $k < 30$ and the averaged values $\tilde{P}_{A(av)}(k)$ are shown for larger block lengths. We increase the gap $\Delta k$ from $\Delta k = 3$ at $k = 30$-$50$ to several tens for the largest observed blocks. The slope of the dashed line describing the asymptotic dependence $\sim k^{-3/2}$ is very close to the slope of the calculated curves and of the marker sequences at quite small $k$. The maximum deviation from these initial linear dependences is observed at the block lengths $k \approx k^* = (L/a)^2$ shown by the dotted lines, for which the second absorbing plane becomes reachable for a chain trajectory. At $k > k^*$, the numbers of trajectories in the shape of loops and bridges become approximately equal o each other, where "loops" and "bridges" are trajectories absorbed at the planes $x = 0$ and $x = L$, respectively [15].

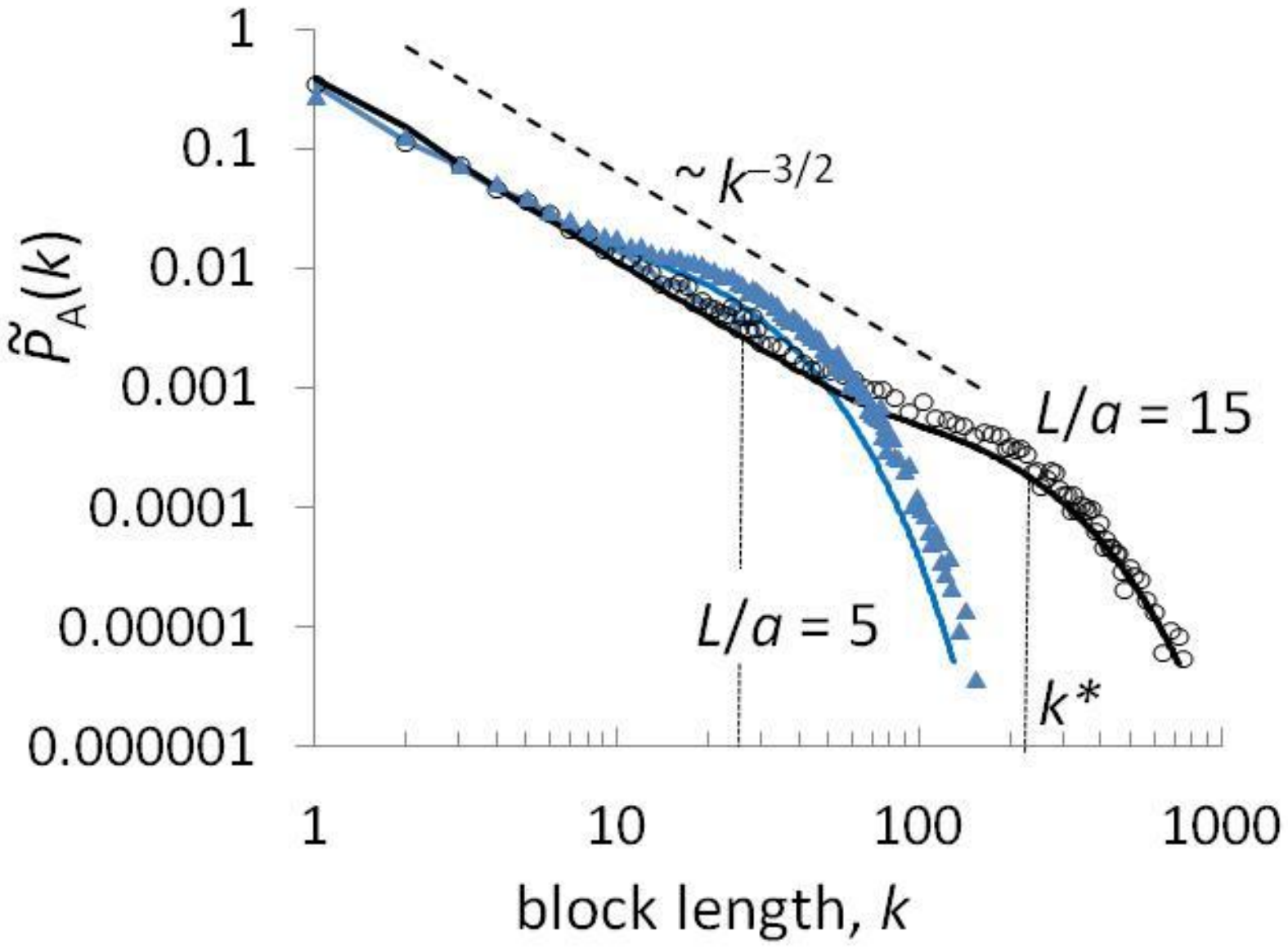

Figure 2. Block length distributions $\tilde{P}_A(k)$ for blocks in the lamellae of thickness $L = L_A = 5a$ (blue curve and markers) and $15a$ (black curve and markers). The curves are calculated from Eq. (23) at $x_0 = 0.5a$. The markers describe the results of the computer simulations. The line describing the dependence $\sim k^{-3/2}$ is plotted to guide the eye. The vertical dotted lines correspond to the values of $k^* = (L/a)^2$.

In figure 3, the number average block size $\bar{k}_n = \bar{k}_{n\mathrm{A}}$ and the dispersity $Đ = Đ_\mathrm{A}$ that characterizes the relative width of the distribution $\tilde{P}_\mathrm{A}(k)$ are presented dependently on the lamellar thickness $L = L_\mathrm{A}$. These values are calculated for the discrete distributions $\tilde{P}_\mathrm{A}(k)$ as

$$\bar{k}_{n\alpha} = \sum_{k=1}^{\infty} k\tilde{P}_\alpha(k), \quad \bar{k}_{w\alpha} = \frac{1}{\bar{k}_{n\alpha}} \sum_{k=1}^{\infty} k^2 \tilde{P}_\alpha(k), \quad Đ_\alpha = \frac{\bar{k}_w}{\bar{k}_n} \tag{24}$$

The dependences of $\bar{k}_n$ and $Đ$ on the lamella thickness are practically linear in the numerical calculations. For the simulation data, the deviation from the linear law is observed at $L/a = 15$, where the presence of chain ends probably influences the distribution width.

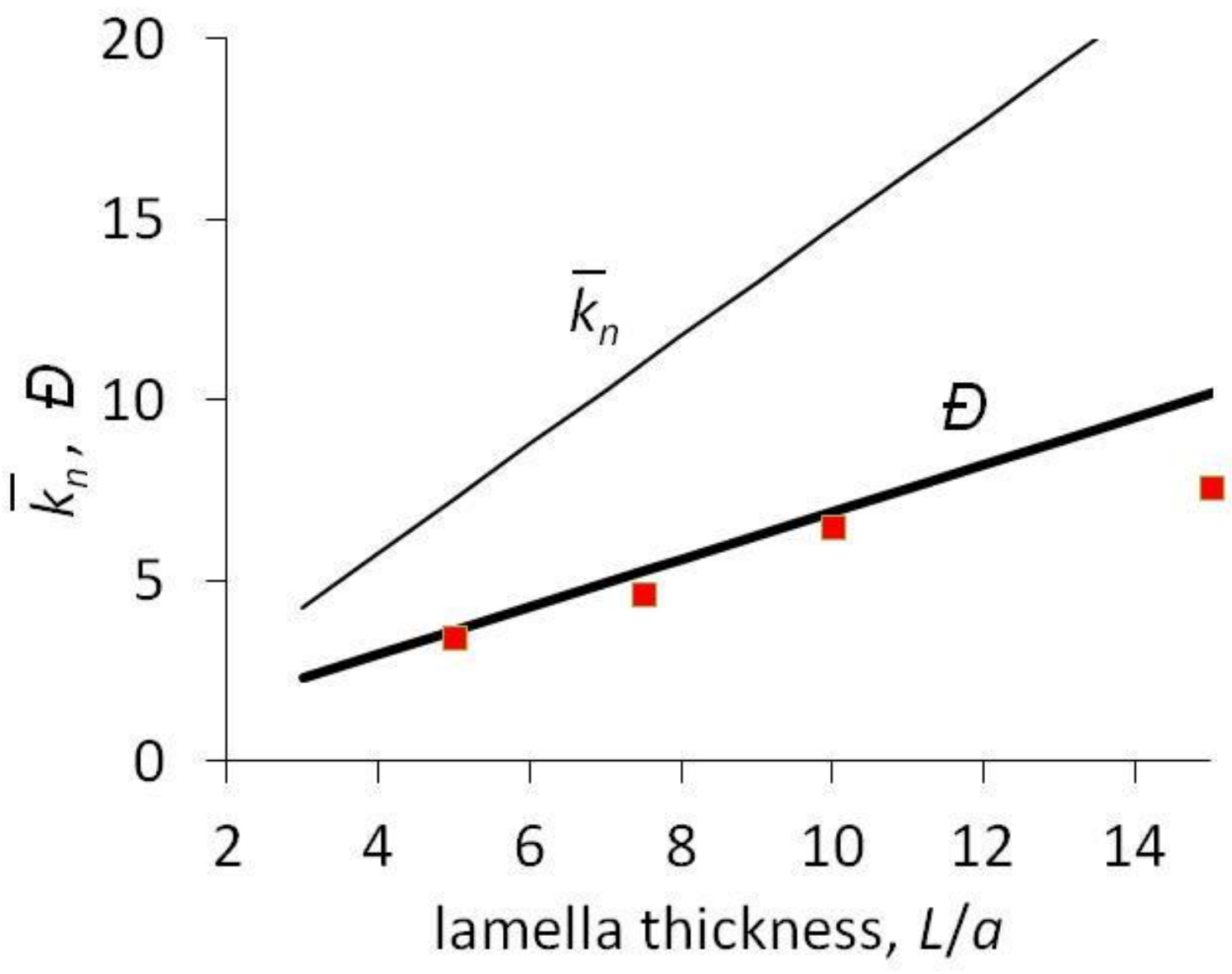

Figure 3. Number average block length $\bar{k}_n = \bar{k}_{n\mathrm{A}}$ (thin solid curve) and dispersity $Đ = Đ_\mathrm{A}$ (thick solid curve and markers) for blocks in the lamella vs lamellar thickness $L = L_\mathrm{A}$. Solid curves and markers describe the results of the numerical calculations and the computer simulations, respectively.

### *3.2 Hexagonally arranged cylinders*

Blocks of type A are located in the hexagonally arranged cylinders of radius $R$ surrounded by the continuous B domain (figure 1b). The distance between the axes of nearest cylinders is equal to $d$. The analytical solution of Eq. (1) for the Green function $G_\mathrm{A}(\mathbf{r},k|\mathbf{r}_0)$ in a cylinder is well known [31], however, the solution for an infinite domain with excluded cylinders has not been reported yet. Any block begins near a cylinder surface that crosses a polymer chain. Any A block is localized within one cylinder. For a B block with the beginning near a certain cylinder, we approximate the presence of other cylinders by a coaxial cylindrical surface of radius $r_\mathrm{out}$ (figure 4).

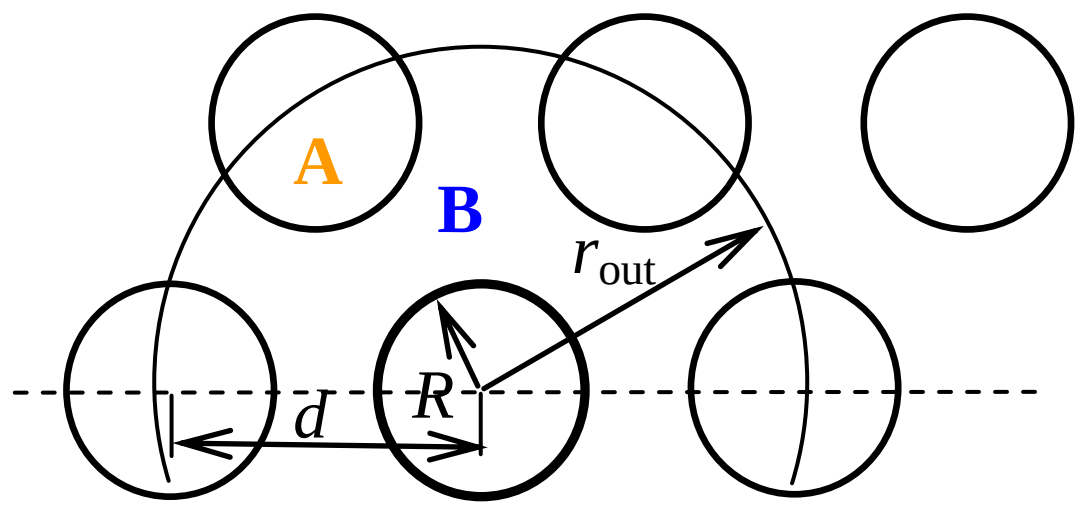

Figure 4. The cross-section of the melt with hexagonally arranged cylinders of radius $R$ with A blocks surrounded by the external domain with B blocks. The distance between the axes of neighboring cylinders is equal to $d$ (the structure period). For the B blocks beginning near the central cylinder, the presence of the other cylinders is described in the theoretical model by the cylindrical surface of radius $r_{out}$.

The three-dimensional Green function of $\alpha$ sequences can be represented as a product $G_\alpha(\mathbf{r},k\,|\,\mathbf{r}_0) = G_{\alpha 2}(r,\varphi,k\,|\,r_0,\varphi_0)G_0(z,k\,|\,z_0)$, where the Green function $G_{\alpha 2}(r,\varphi,k\,|\,r_0,\varphi_0)$ obeys the equation

$$\frac{\partial G_{\alpha 2}}{\partial k} = \frac{a^2}{6}\left(\frac{1}{r}\frac{\partial}{\partial r}\left(r\frac{\partial G_{\alpha 2}}{\partial r}\right) + \frac{1}{r^2}\frac{\partial^2 G_{\alpha 2}}{\partial \varphi^2}\right) \tag{25}$$

The function $G_{A2}$ is defined for a circle at $0 \le r \le R$ and $0 \le \varphi \le 2\pi$, the boundary condition is $G_{A2}(R,\varphi,k\,|\,r_0,\varphi_0) = 0$. The function $G_{B2}$ is defined for a ring at $R \le r \le r_{out}$ and $0 \le \varphi \le 2\pi$, the boundary condition is $G_{B2}(R,\varphi,k\,|\,r_0,\varphi_0) = G_{B2}(r_{out},\varphi,k\,|\,r_0,\varphi_0) = 0$. The initial condition for the both functions is $G_{\alpha 2}\big|_{k=0} = \frac{1}{r}\delta(r-r_0)\delta(\varphi-\varphi_0)$. Since the block length distribution does not depend on the choice of the orientations of the coordinate axes, the convenient value $\varphi_0 = 0$ can be used.

The Green function $G_{A2}$ for a circle has the form

$$G_{A2} = \frac{1}{\pi R^2}\sum_{j=1}^{\infty}\sum_{n=0}^{\infty}\frac{b_n \cos n\varphi}{\left[J_n'\left(\mu_j^{(n)}\right)\right]^2} J_n\left(\mu_j^{(n)}\frac{r_0}{R}\right) J_n\left(\mu_j^{(n)}\frac{r}{R}\right)\exp\left(-\frac{a^2(\mu_j^{(n)})^2}{6R^2}k\right) \tag{26}$$

where $J_n$ is the Bessel function of the $n$-th order, $b_n = 1$ for $n = 0$ and $b_n = 2$ for $n \ne 0$, $\mu_j^{(n)}$ is the $j$-th positive solution of the equation $J_n(\mu) = 0$. The values of these solutions for the Bessel function of the zeroth order are $\mu_1^{(0)} = 2.40$, $\mu_2^{(0)} = 5.52$, $\mu_3^{(0)} = 8.65$, $\mu_4^{(0)} = 11.79$, $\mu_{j+1}^{(0)} - \mu_j^{(0)} \approx \pi$.

The block length distribution $P_A(k)$ for blocks in cylinders can be found using Eqs. (5) and (24) as $P_A(k) = -\frac{\partial}{\partial k}\int_0^R dr r \int_0^{2\pi} d\varphi G_{A2}$. Since the terms with $n > 0$ are equal to zero, the distribution $P_A(k)$ can be written as

$$P_A(k) = \frac{1}{3\tilde{R}^2}\sum_{j=1}^{\infty}\frac{(\mu_j^{(0)})^2}{\left[J_0'\left(\mu_j^{(0)}\right)\right]^2} J_0\left(\mu_j^{(0)}\frac{\tilde{r}_0}{\tilde{R}}\right) f_0(\mu_j^{(0)}) \exp\left(-\frac{(\mu_j^{(0)})^2}{6\tilde{R}^2}k\right), \qquad (27)$$

where $f_0(\mu) = \int_0^1 dx\, x J_0(\mu x)$, $\tilde{R} = R/a$, $\tilde{r}_0 = r_0/a$. Using the relations between the Bessel functions of the 0th and the 1st order, $P_A(k)$ can be written as

$$P_A(k) = \frac{1}{3\tilde{R}^2}\sum_{j=1}^{\infty}\frac{\mu_j^{(0)}}{J_1\left(\mu_j^{(0)}\right)} J_0\left(\mu_j^{(0)}\frac{\tilde{r}_0}{\tilde{R}}\right) \exp\left(-\frac{(\mu_j^{(0)})^2}{6\tilde{R}^2}k\right) \qquad (28)$$

Then, the probability distribution in a discrete form is

$$\tilde{P}_A(k) = 2\sum_{j=1}^{\infty}\frac{J_0\left(\mu_j^{(0)}\frac{\tilde{r}_0}{\tilde{R}}\right)}{\mu_j^{(0)} J_1\left(\mu_j^{(0)}\right)}\left(\exp\left(-\frac{(\mu_j^{(0)})^2}{6\tilde{R}^2}(k-1)\right) - \exp\left(-\frac{(\mu_j^{(0)})^2}{6\tilde{R}^2}k\right)\right). \qquad (29)$$

The Bessel functions were calculated numerically using the necessary numbers of terms in the sums $J_0(x) = \sum_{n=0}^{\infty}\frac{(-1)^n}{(n!)^2}\left(\frac{x}{2}\right)^{2n}$, $J_1(x) = \left(\frac{x}{2}\right)\sum_{n=0}^{\infty}\frac{(-1)^n}{(n+1)!n!}\left(\frac{x}{2}\right)^{2n}$ for $x < 37$. For larger arguments, the asymptotic form $J_0(x) = \sqrt{\frac{2}{\pi x}}\left(\cos(x-\pi/4) + \frac{1}{8x}\sin(x-\pi/4)\right)$ was taken, $J_1(x) = -J_0'(x)$.

The Green function $G_{B2}$ for B blocks in a ring has the form

$$G_{B2} = \frac{\pi}{4R^2}\sum_{j=1}^{\infty}\sum_{n=0}^{\infty}\frac{b_n(\tilde{\mu}_j^{(n)})^2 J_n^2\left(\tilde{\mu}_j^{(n)}c_{out}\right)\cos n\varphi}{J_n^2\left(\tilde{\mu}_j^{(n)}\right) - J_n^2\left(\tilde{\mu}_j^{(n)}c_{out}\right)} f_n\left(\tilde{\mu}_j^{(n)}, \frac{\tilde{r}_0}{\tilde{R}}\right) f_n\left(\tilde{\mu}_j^{(n)}, \frac{\tilde{r}}{\tilde{R}}\right)\exp\left(-\frac{(\tilde{\mu}_j^{(n)})^2}{6R^2}k\right), \qquad (30)$$

where $f_n(\mu, c) = N_n(\mu)J_n(\mu c) - J_n(\mu)N_n(\mu c)$, $N_n$ is the Bessel function of the second kind (the Neumann function) of the $n$-th order, $\tilde{\mu}_j^{(n)}$ is the $j$-th positive solution of the equation $f_n(\mu, c_{out}) = 0$, $c_{out} = r_{out}/R$.

Using equations (5) and (30), the probability distribution $P_B(k)$ is calculated as $P_B(k) = -\frac{\partial}{\partial k}\int_R^{r_{out}} dr r \int_0^{2\pi} d\varphi G_{B2}$:

$$P_B(k) = \frac{\pi^2}{12\tilde{R}^2} \sum_{j=1}^{\infty} \frac{(\tilde{\mu}_j^{(0)})^4 J_0^2\left(\tilde{\mu}_j^{(0)} c_{out}\right)}{J_0^2\left(\tilde{\mu}_j^{(0)}\right) - J_0^2\left(\tilde{\mu}_j^{(0)} c_{out}\right)} f_0\left(\tilde{\mu}_j^{(0)}, \frac{\tilde{r}_0}{\tilde{R}}\right) f_{\rm B}(\tilde{\mu}_j^{(0)}, c_{out}) \exp\left(-\frac{(\tilde{\mu}_j^{(0)})^2}{6\tilde{R}^2} k\right), \quad (31)$$

where $f_{\rm B}(\mu, c) = \int_1^c dx x f_0(\mu, x)$ . Here, the Neumann function of the zeroth order can be calculated as [35]

$$N_0(x) = \frac{2}{\pi}\left(\gamma + \log\left(\frac{x}{2}\right)\right) J_0(x) - f_{\rm h}(x), \quad f_{\rm h}(x) = \frac{2}{\pi} \sum_{m=1}^{\infty} \frac{(-1)^m}{(m!)^2} \left(\frac{x}{2}\right)^{2m} h_m,$$

where $\gamma = 0.5772157\ldots$ denotes the Euler constant, $h_m = 1^{-1} + 2^{-1} + \ldots + m^{-1}$. The function $f_{\rm B}(\mu, c)$ can be rewritten as

$$f_{\rm B}(\mu, c) = -\frac{2}{\pi\mu} J_0(\mu)\left\{ J_1(\mu c) c \ln c + \frac{1}{\mu}\left(J_0(\mu c) - J_0(\mu)\right)\right\} - \frac{2}{\pi\mu}\left(c J_1(\mu c) - J_1(\mu)\right) f_{\rm h}(\mu) + \frac{1}{\pi} J_0(\mu) f_{\rm h1}(\mu, c), \quad (32)$$

where $f_{\rm h1}(\mu, c) = \sum_{m=1}^{\infty} \frac{(-1)^m h_m}{(m!)^2 (m+1)} \left(\frac{x}{2}\right)^{2m} \left(c_{\rm out}^{2(m+1)} - 1\right)$. The values of $f_{\rm B}(\mu, c)$ were calculated using the expression (32) at $\mu c < 37$, at larger arguments the function $f_0(\mu, c)$ was integrated numerically.

The probability distribution $P_{\rm B}(k)$ in a discrete form is

$$\tilde{P}_B(k) = \frac{\pi^2}{2} \sum_{j=1}^{\infty} \frac{(\tilde{\mu}_j^{(0)})^2 J_0^2\left(\tilde{\mu}_j^{(0)} c_{out}\right)}{J_0^2\left(\tilde{\mu}_j^{(0)}\right) - J_0^2\left(\tilde{\mu}_j^{(0)} c_{out}\right)} f_0\left(\tilde{\mu}_j^{(0)}, \frac{\tilde{r}_0}{\tilde{R}}\right) f_{\rm B}(\tilde{\mu}_j^{(0)}, c_{out}) \left(\exp\left(-\frac{(\tilde{\mu}_j^{(0)})^2}{6\tilde{R}^2}(k-1)\right) - \exp\left(-\frac{(\tilde{\mu}_j^{(0)})^2}{6\tilde{R}^2} k\right)\right) \quad (33)$$

The results for the probability distributions $\tilde{P}_{\rm A}(k)$ for blocks in the cylinder of radius $R = 4a$ and $\tilde{P}_{\rm B}(k)$ for blocks in the external domain are presented in figure 5. The external domain in the computer simulations encompasses the cylinders of radius $R = 4a$ packed into the hexagonal structure with a period $d = 15a$. In the numerical calculations, the probability distribution $\tilde{P}_{\rm B}(k)$ is presented for two cylindrical layers between $R = 4a$ and two outer radii $r_{\rm out} = 15a$ (thin blue curve) and $r_{\rm out} = 13.1a$ (thick blue curve). The second curve provides the best fit for the slope of the dependence of $\tilde{P}_{\rm B}(k)$ on $k$ in the inset graph.

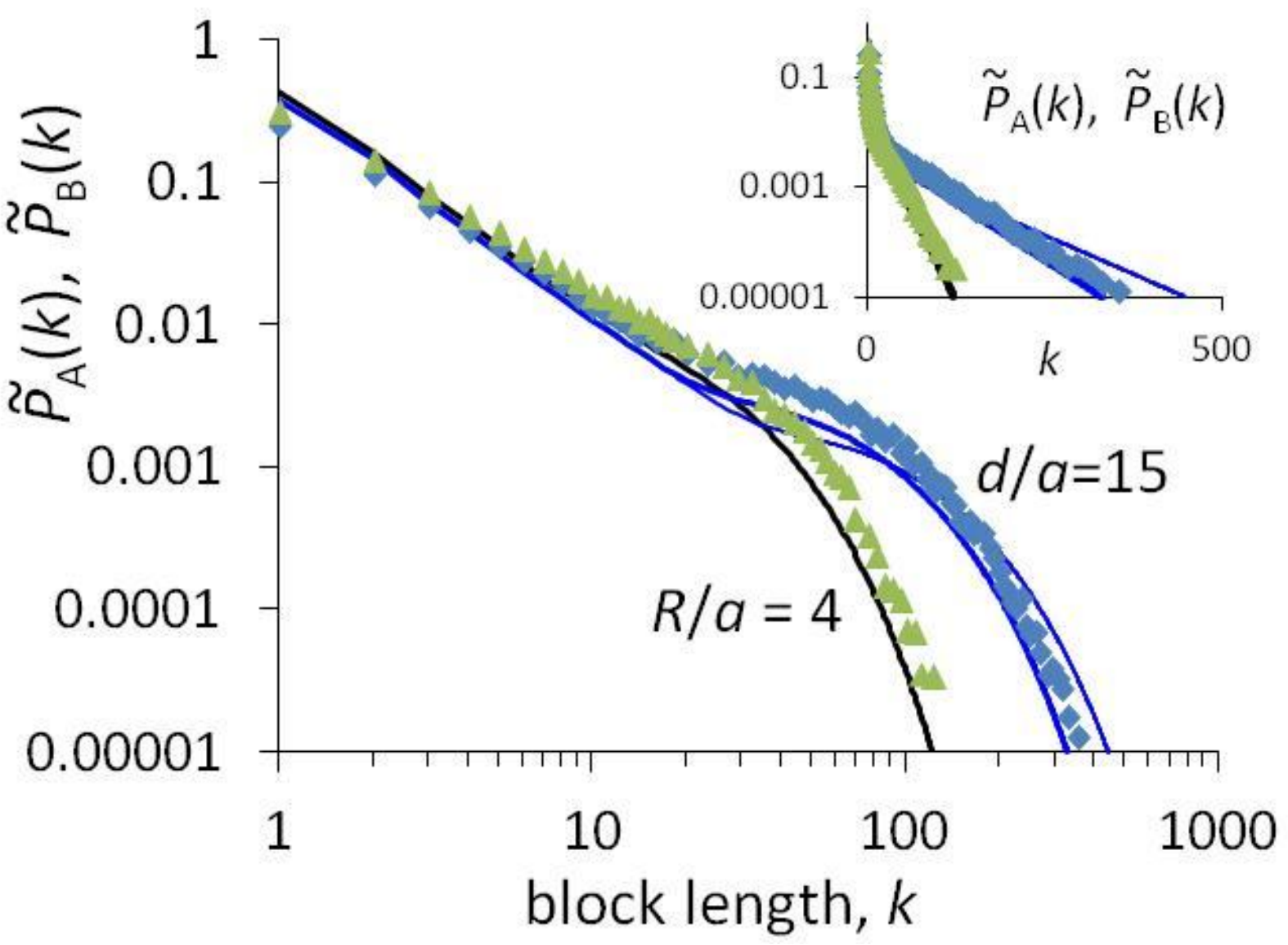

Figure 5. Block length distributions $\tilde{P}_A(k)$ for A blocks in the cylinder of radius $R = 4a$ and $\tilde{P}_B(k)$ for B blocks in the external domain. Markers represent the results of the computer simulations for the cylinder (triangles) and for the external domain (rhombuses) of the hexagonal structure with the period $d = 15a$. Solid curves describe the results of the numerical calculations for the cylinder (thick black curve, $R - r_0 = 0.5a$) and for the cylindrical layer with $R/a = 4$, $r_0 - R = 0.5a$, $r_{out}/a = 15$ (thin blue curve) and 13.1 (thick blue curve). The inset represents the same dependences in the plot with the logarithmic scale only at the vertical axis.

In figure 6, the number average block length $\bar{k}_{nA}$ and the dispersity $Đ_A$ for A blocks in the cylinders are plotted dependently on the cylinder radius $R$. The dependences are practically linear both in the theory and computer simulations. In figure 7, the number average block length $\bar{k}_{nB}$ and the dispersity $Đ_B$ for B blocks in the external domain are plotted dependently on the structure period $d$ in the computer simulations or on the outer radius $r_{out}$ in the numerical calculations at the fixed cylinder radius $R$. The same parameters are presented in figure 8 depending on the cylinder radius $R$ at the fixed sizes of the external domain.

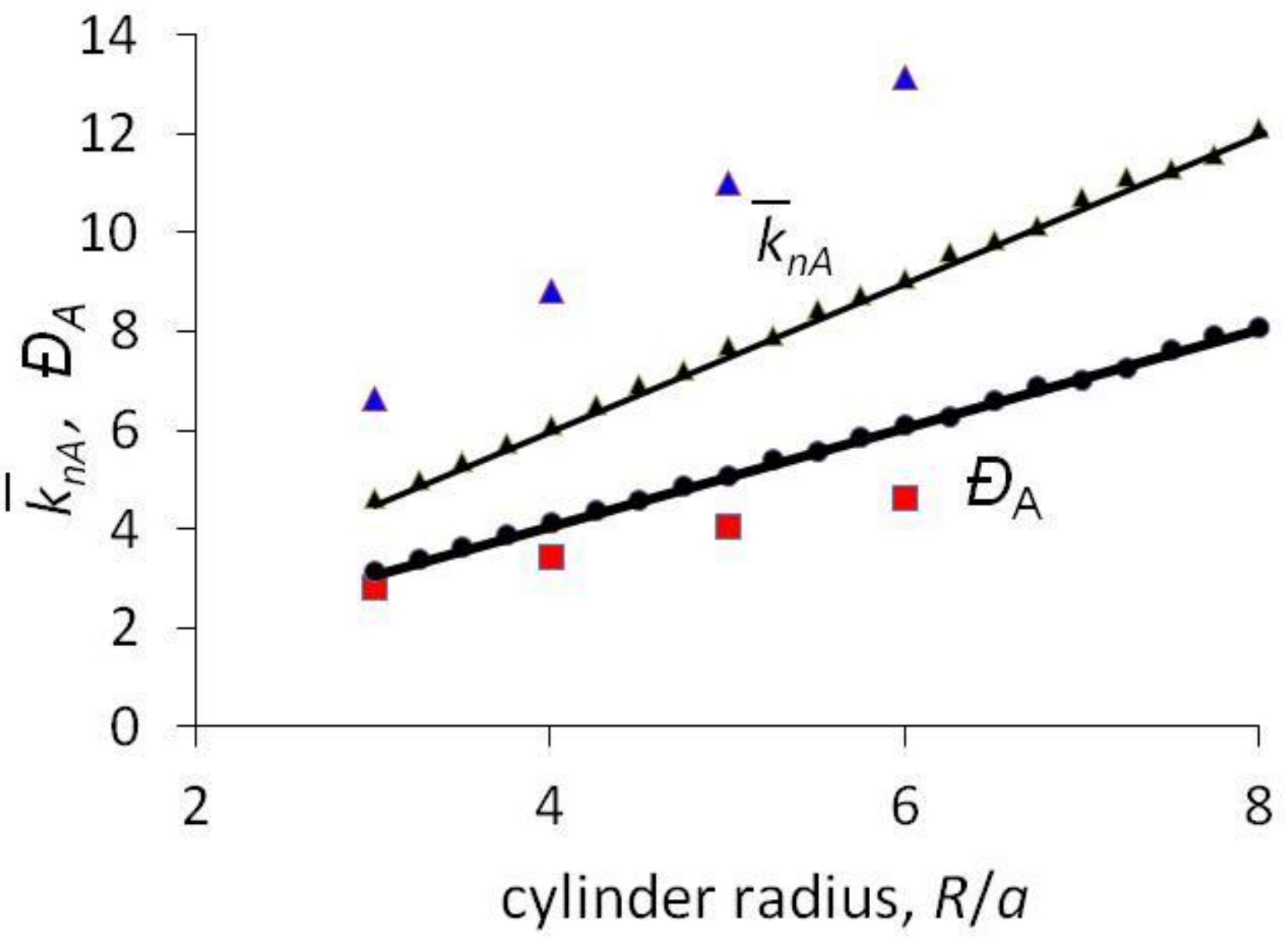

Figure 6. Number average block length $\bar{k}_{n\mathrm{A}}$ and dispersity *Đ* of blocks A in the cylinders vs cylinder radius *R*/*a*. Solid curves with markers describe the results of the numerical calculations for $\bar{k}_{n\mathrm{A}}$ (thin curve with triangles) and *Đ* (thick curve with circles). The separate markers describe the results of the computer simulations for $\bar{k}_{n\mathrm{A}}$ (blue triangles) and *Đ* (red squares).

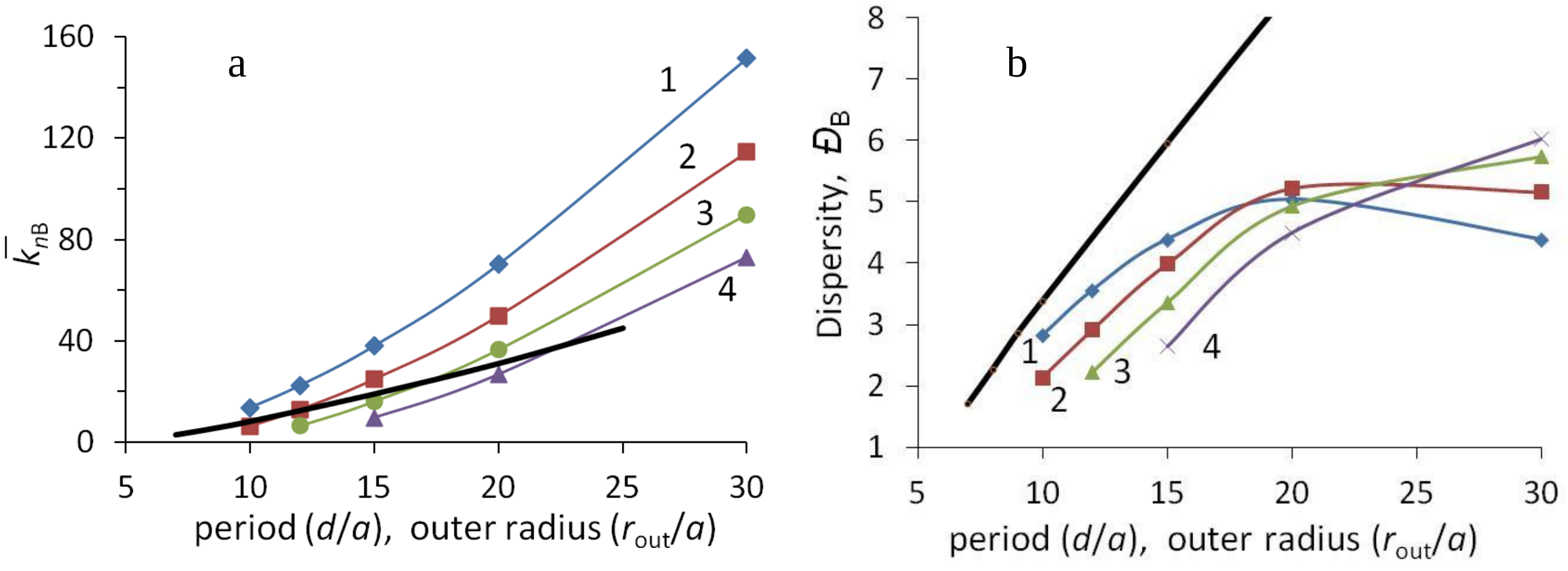

Figure 7. Dependence of the number average block length $\bar{k}_{n\mathrm{B}}$ (a) and dispersity $Đ_{\mathrm{B}}$ (b) of B blocks in the external domain on spatial period *d*/*a* (thin curves with markers) at the values of cylinder radius *R*/*a* = 3 (1), 4 (2), 5 (3), and 6 (4) in the computer simulations and on outer radius $r_{\mathrm{out}}$ (thick black curve) at the radius *R* = 5*a* in the numerical calculations.

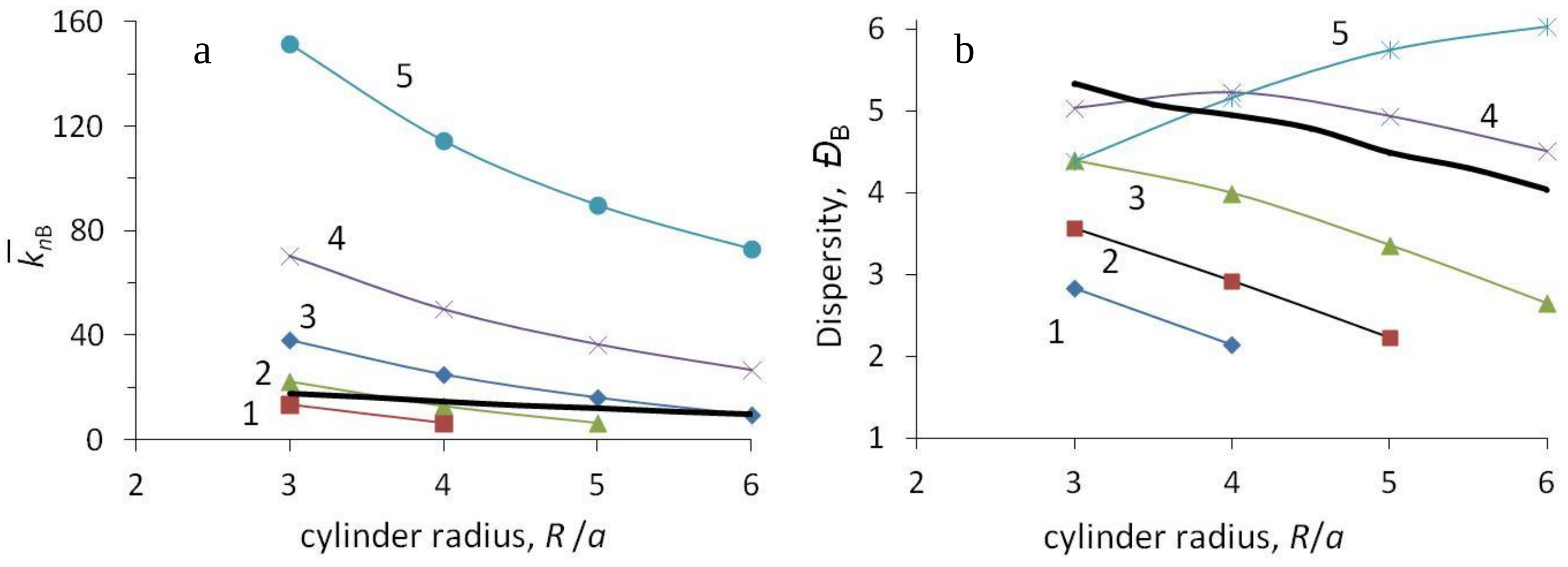

Figure 8. Dependence of the number average block length $\bar{k}_{n\mathrm{B}}$ (a) and dispersity $Đ_\mathrm{B}$ (b) of B blocks in the external domain on cylinder radius $R/a$ in the computer simulations (thin curves with markers) at the values of spatial period $d/a = 10$ (1), 12 (2), 15 (3), 20 (4), 30 (5) and in the numerical calculations (thick black curves) at the outer radius $r_\mathrm{out} = 12a$.

*3.2 Balls packed in a body-centered cubic lattice*

Balls of radius $R_\mathrm{s}$ are arranged in a body-centered cubic lattice, those balls consist of A blocks and the surrounding domain consists of B blocks (figure 1c). The distance between the centers of nearest balls is equal to $d_\mathrm{s}$. The solution of Eq. (1) for sequences of A monomer units in a ball can be written in a usual way [31], whereas the form of the solution for an infinite region with excluded balls is not known. For a B block beginning near a certain ball, the presence of other balls is approximated by an absorbing spherical surface of radius $r_\mathrm{out}$ with the same center.

The Green function of $\alpha$-type sequences obeys Eq. (1) that can be written in the spherical coordinates for the function $G_\alpha(r,\theta,\varphi,k|r_0,\theta_0,\varphi_0)$ as

$$\frac{\partial G_\alpha}{\partial k} = \frac{a^2}{6r^2}\left(\frac{\partial}{\partial r}\left(r^2\frac{\partial G_\alpha}{\partial r}\right) + \frac{1}{\sin\theta}\frac{\partial}{\partial\theta}\left(\sin\theta\frac{\partial G_\alpha}{\partial\theta}\right) + \frac{1}{\sin^2\theta}\frac{\partial^2 G_\alpha}{\partial\varphi^2}\right) \quad (34)$$

The function $G_\mathrm{A}$ is defined for a ball at $0 \le r \le R_\mathrm{s}$, $0 \le \theta \le \pi$, and $0 \le \varphi \le 2\pi$; the boundary condition is $G_\mathrm{A}|_{r=R_\mathrm{S}} = 0$. The function $G_\mathrm{B}$ is defined for a spherical layer at $R_\mathrm{s} \le r \le r_{out}$, $0 \le \theta \le \pi$, and $0 \le \varphi \le 2\pi$; the boundary condition is $G_\mathrm{B}|_{r=R_\mathrm{S}} = G_\mathrm{B}|_{r=r_{out}} = 0$. The initial condition for the both functions is $G_\alpha|_{k=0} = \frac{1}{r^2}\delta(r-r_0)\delta(\theta-\theta_0)\delta(\varphi-\varphi_0)$. The convenient orientations of the coordinate axes can be chosen to provide the initial values $\theta_0 = 0$, $\varphi_0 = 0$.

The Green function $G_\mathrm{A}$ for the ball has the form

$$G_{\rm A} = \frac{1}{2\pi R_{\rm s}^2 \sqrt{r r_0}} \sum_{j=1}^{\infty} \sum_{n=0}^{\infty} \frac{(2n+1) P_n(\cos\theta)}{\left( J'_{n+1/2}\left( \mu_j^{(n+1/2)} \right) \right)^2} J_{n+1/2}\left( \mu_j^{(n+1/2)} \frac{r_0}{R_{\rm s}} \right) J_{n+1/2}\left( \mu_j^{(n+1/2)} \frac{r}{R_{\rm s}} \right)$$
$$\times \exp\left( -\frac{1}{6} \left( \frac{\mu_j^{(n+1/2)}}{\tilde{R}_{\rm s}} \right)^2 k \right) \tag{35}$$

where $P_n(x)$ is the Legendre polynomial, $\mu_j^{(n+1/2)}$ is the $j$-th positive solution of the equation $J_{n+1/2}(\mu) = 0$, $\tilde{R}_{\rm s} = R_{\rm s}/a$. Since $J_{1/2}(\mu) = \sqrt{\frac{2}{\pi\mu}} \sin\mu$, these solutions at $n = 0$ are $\mu_j^{(1/2)} = \pi j$.

The block length distribution $P_{\rm A}(k)$ for blocks in the ball can be found as $P_{\rm A}(k) = -\frac{\partial}{\partial k} \int_0^{R_{\rm s}} dr r^2 \int_0^{\pi} d\theta \sin\theta \int_0^{2\pi} d\varphi G_{\rm A}$, where only the terms with $n = 0$ give non-zero contributions:

$$P_{\rm A}(k) = \frac{\pi}{3\tilde{R}_{\rm s} \tilde{r}_0} \sum_{j=1}^{\infty} j(-1)^{j+1} \sin\left( j\pi \frac{r_0}{R_{\rm s}} \right) \exp\left( -\frac{j^2 \pi^2}{6\tilde{R}_{\rm s}^2} k \right) \tag{36}$$

In a discrete form,

$$\tilde{P}_{\rm A}(k) = \frac{2\tilde{R}_{\rm s}}{\pi \tilde{r}_0} \sum_{j=1}^{\infty} \frac{(-1)^{j+1}}{j} \sin\left( j\pi \frac{r_0}{R_{\rm s}} \right) \left( \exp\left( -\frac{j^2\pi^2}{6\tilde{R}_{\rm s}^2} (k-1) \right) - \exp\left( -\frac{j^2\pi^2}{6\tilde{R}_{\rm s}^2} k \right) \right) \tag{37}$$

The solution of Eq. (34) for the function $G_{\rm B}$ in a spherical layer at $\theta_0 = 0$, $\varphi_0 = 0$ can be written in the form

$$G_{\rm B} = \frac{\pi}{8 R_{\rm s}^2 \sqrt{r r_0}} \sum_{j=1}^{\infty} \sum_{n=0}^{\infty} \frac{(2n+1) P_n(\cos\theta) (\tilde{\mu}_j^{(n+1/2)})^2 J_{n+1/2}^2 (\tilde{\mu}_j^{(n+1/2)} \frac{r_{\rm out}}{R_s})}{J_{n+1/2}^2 (\tilde{\mu}_j^{(n+1/2)}) - J_{n+1/2}^2 (\tilde{\mu}_j^{(n+1/2)} \frac{r_{\rm out}}{R_s})}$$
$$\times f_{n+1/2}(\tilde{\mu}_j^{(n+1/2)}, \frac{r_0}{R_s}) f_{n+1/2}(\tilde{\mu}_j^{(n+1/2)}, \frac{r}{R_s}) \exp\left( -\frac{1}{6} \frac{(\tilde{\mu}_j^{(n+1/2)})^2}{\tilde{R}_{\rm s}^2} k \right) \tag{38}$$

where $f_{n+1/2}(\mu, c) = N_{n+1/2}(\mu) J_{n+1/2}(\mu c) - J_{n+1/2}(\mu) N_{n+1/2}(\mu c)$, $\tilde{\mu}_j^{(n+1/2)}$ is the $j$-th positive solution of the equation $f_{n+1/2}(\mu, c_{\rm out}) = 0$, $c_{\rm out} = r_{\rm out}/R_s$ .

The block length distribution $P_{\rm B}(k)$ can be calculated as $P_{\rm B}(k) = -\frac{\partial}{\partial k} \int_{R_s}^{r_{\rm out}} dr r^2 \int_0^{\pi} d\theta \sin\theta \int_0^{2\pi} d\varphi G_{\rm B}$. The integrals of $P_n(\cos\theta)$ over $\theta$ are not equal to zero only for $n = 0$, therefore, only the terms with $P_0(\cos\theta) = 1$ should be taken. The equation $f_{1/2}(\mu, c_{\text{out}}) = 0$ is equivalent to $\sin\mu(c_{\text{out}} - 1) = 0$. Introducing the notations $\Delta R = r_{\text{out}} - R_s$, $\Delta r_0 = r_0 - R_s$, $\Delta\tilde{R} = \Delta R/a$, the block length distribution can be written in the form

$$P_B(k) = \frac{\pi R_s a^2}{3 r_0 \Delta R^2} \sum_{j=1}^{\infty} j \sin\left(\pi j \frac{\Delta r_0}{\Delta R}\right)\left(1 - (-1)^j \frac{r_{\text{out}}}{R_s}\right) \exp\left(-\frac{(\pi j)^2}{6\Delta\tilde{R}^2} k\right), \tag{39}$$

where the value $\tilde{\mu}_j^{(1/2)} = \dfrac{\pi j R_s}{\Delta R}$ is used.

In a discrete form,

$$P_B(k) = \frac{2R_s}{\pi r_0} \sum_{j=1}^{\infty} \frac{1}{j} \sin\left(\pi j \frac{\Delta r_0}{\Delta R}\right)\left(1 - (-1)^j \frac{r_{\text{out}}}{R_s}\right)\left(\exp\left(-\frac{(\pi j)^2}{6\Delta\tilde{R}^2}(k-1)\right) - \exp\left(-\frac{1}{6}\frac{(\pi j)^2}{\Delta\tilde{R}^2} k\right)\right) \tag{40}$$

The block length distributions $\tilde{P}_A(k)$ for A blocks in the cylinder and in the ball of the same radius $5a$ are presented in figure 9. The distributions found in the computer simulations are very close to each other at $k < (R/a)^2$, whereas shorter blocks prevail in the ball at larger $k$ in comparison with the cylinder. The block length distributions $\tilde{P}_B(k)$ of B blocks in the external domains of different sizes are plotted in figure 10 at $R_s = 5a$. The solid curves are plotted for the numerical data at the values of $r_{\text{out}}$ that provide the best fit for the slope of the dependences obtained in the computer simulations at large $k$.

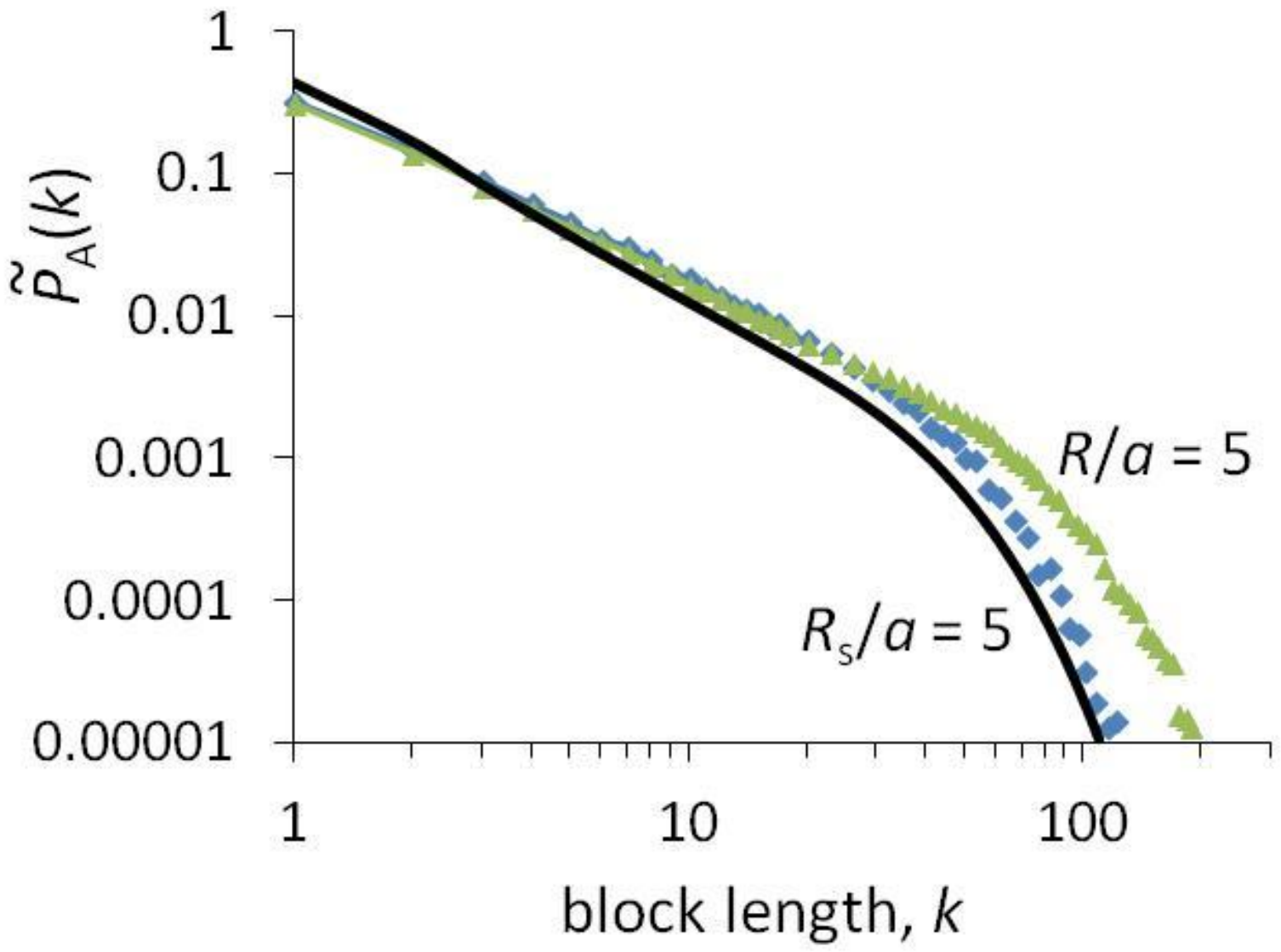

Figure 9. Block length distributions $\tilde{P}_A(k)$ of A blocks in the cylinder of radius $R = 5a$ (green triangles) and in the ball of radius $R_s = 5a$ (blue rhombuses and solid curve). The markers represent the results of the computer simulations, the solid curve describes the numerical calculations, $R - r_0 = 0.5a$.

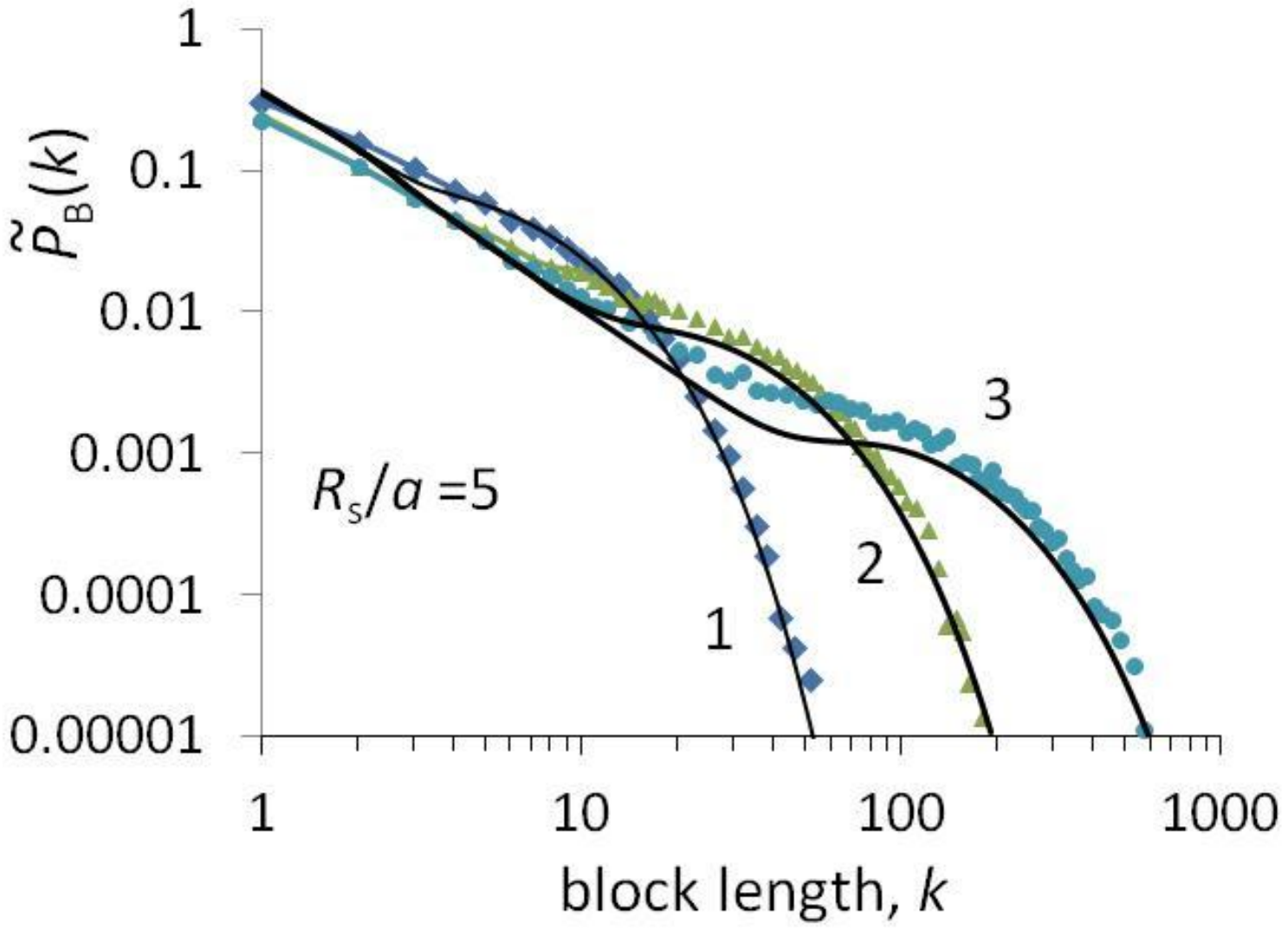

Figure 10. Block length distribution $\tilde{P}_B(k)$ of B blocks in the external domain surrounding a ball of radius $R_s = 5a$ at $d_s/a = 12$ (1), 15 (2), and 20 (3) in the computer simulations (markers) and at $r_{out}/a = 8$ (1), 11.5 (2), and 18 (3) in the numerical calculations (black solid curves, $r_0 - R = 0.5a$).

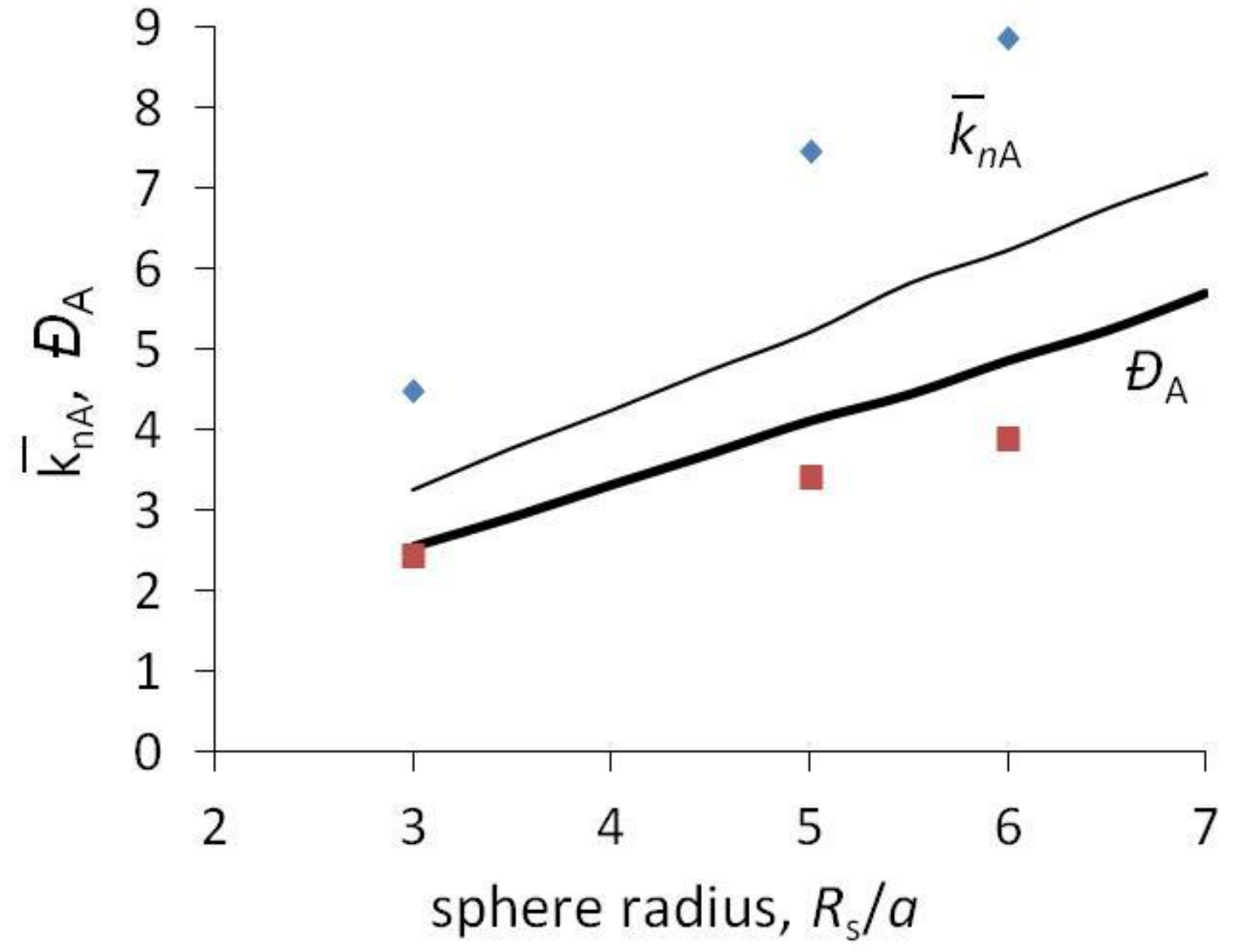

Figure 11. Number average block length $\bar{k}_{nA}$ and dispersity $Đ_A$ of blocks A in a ball vs ball radius $R_s/a$. Solid curves describe the results of the numerical calculations for $\bar{k}_{nA}$ (thin curve) and $Đ$ (thick curve). Markers describe the results of the computer simulations for $\bar{k}_{nA}$ (blue rhombuses) and $Đ$ (red squares).

In figure 11, the number average block length $\bar{k}_{n\mathrm{A}}$ and the dispersity $Đ_\mathrm{A}$ for A blocks in a ball are plotted dependently on its radius $R_\mathrm{s}$, and they grow approximately linearly with the radius as well as for a cylinder. In figures 12 and 13, the dependences of the number average block length $\bar{k}_{n\mathrm{B}}$ and the dispersity $Đ_\mathrm{B}$ for B blocks in the external domain are plotted vs the structure period $d$ (or the outer radius $r_\mathrm{out}$) and the ball radius $R$.

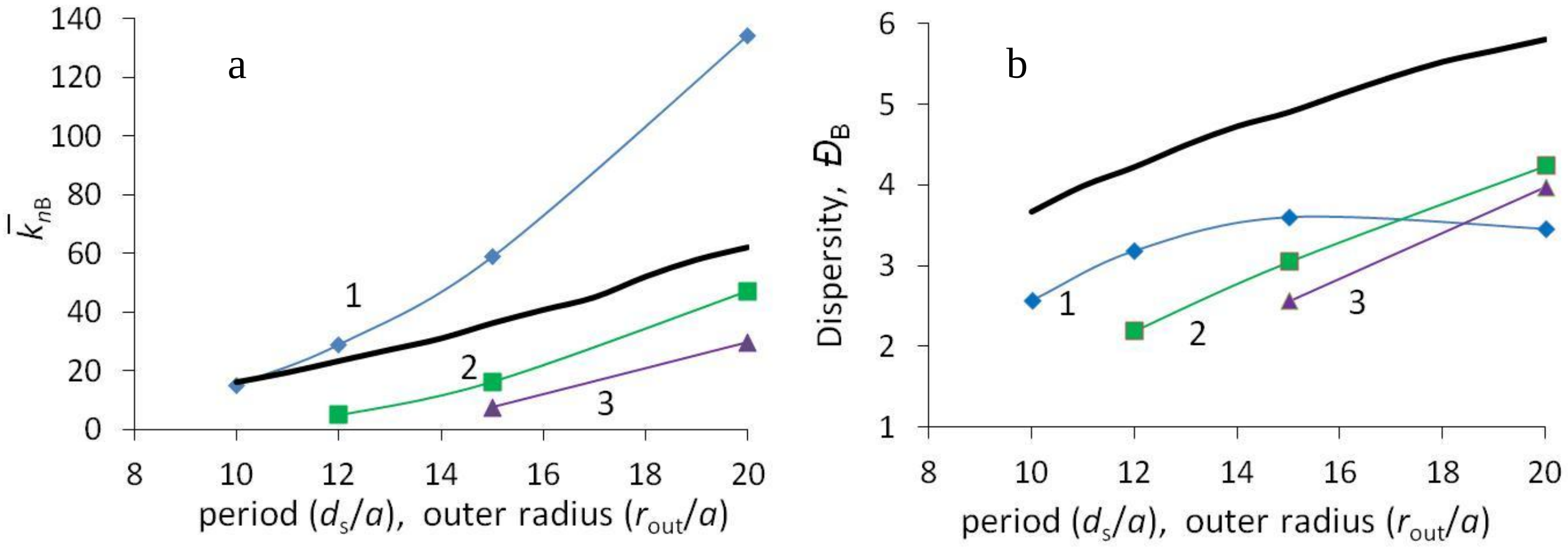

Figure 12. Dependence of the number average block length $\bar{k}_{n\mathrm{B}}$ (a) and dispersity $Đ_\mathrm{B}$ (b) of B blocks in the external domain on spatial period $d_\mathrm{s}/a$ (thin curves with markers) at the values of ball radius $R_\mathrm{s}/a$ = 3 (1), 5 (2), and 6 (3) in the computer simulations and on the outer radius $r_\mathrm{out}$ (thick black curves) at the radius $R_\mathrm{s} = 3a$ in the numerical calculations.

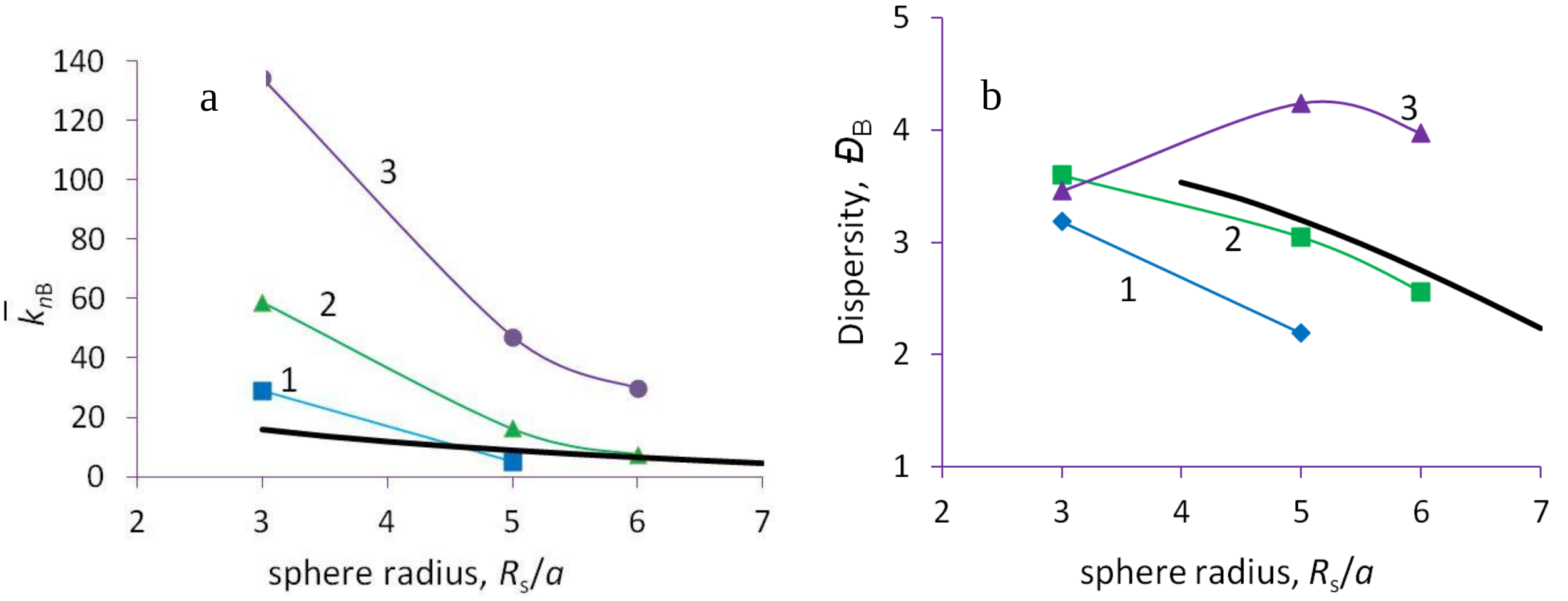

Figure 13. Dependence of the number average block length $\bar{k}_{n\mathrm{B}}$ (a) and dispersity $Đ_\mathrm{B}$ (b) of B blocks in the external domain on ball radius $R_\mathrm{s}/a$ in the computer simulations (thin curves with markers) at the values of spatial period $d_\mathrm{s}/a$ = 12 (1), 15 (2), 20 (3) and on the outer radius $r_\mathrm{out}/a$ in the numerical calculations (thick black curves) at the outer radius $r_\mathrm{out} = 10a$.

*3.3 Statistical characteristics of sequences*

In the previous three subsections the formulae for the block length distributions are derived for three considered structure types (figure 1), and the dependences of the probabilities $\tilde{P}_\alpha(k)$ on the block length $k$ and of the distribution parameters $\bar{k}_{n\alpha}$, $Đ_\alpha$ on the domain sizes are plotted. All the probabilities $\tilde{P}_\alpha(k)$ (figures 2, 5, 9, and 10) obviously obey the asymptotic law $\sim k^{-3/2}$ at the moderate values of block length (the asymptotic line is shown in figure 2). At very large $k$, the probabilities decrease exponentially in the theory and practically by the same way in the computer simulations.

The block length distributions $\tilde{P}_\mathrm{A}(k)$ obtained with the computer simulations and with the numerical calculations at the same values of the domain size (lamella thickness $L_\mathrm{A}$, cylinder radius $R$, or ball radius $R_\mathrm{s}$) are close to each other (figures 2, 5, 9). The distribution for a cylinder is wider than for a ball of the same radius $5a$ (figure 9), since A blocks are bounded in all three dimensions in the ball, whereas there are no boundaries along the cylinder axes. The ratio of the number average values of block lengths for a cylinder and a ball of radius $5a$ is approximately equal to $\bar{k}_{n\mathrm{A(cyl)}}/\bar{k}_{n\mathrm{A(ball)}} \approx 1.48$ both in the numerical calculations and in the computer simulations (figures 6 and 11) despite the values of the lengths $\bar{k}_{n\mathrm{A}}$ themselves are different.

The probability distributions $\tilde{P}_\mathrm{A}(k)$ for lamellae (figure 2) and $\tilde{P}_\mathrm{B}(k)$ for the external regions of cylinders and balls (figures 5 and 10) can demonstrate non-monotonic change of the slope in the double logarithmic scale. The hump at the intermediate values of block length $k$ corresponds to the critical value $k^*$, for which a block with $k > k^*$ can reach an additional absorbing boundary. That value can be estimated as $k^* = (L/a)^2$ for lamellae or $(\Delta R/a)^2$ for cylindrical and spherical layers of thickness $\Delta R$ in the numerical calculations. In the computer simulations, such hump is less pronounced and can be related to $k^* = (\Delta R_\mathrm{min}/a)^2$, where $\Delta R_\mathrm{min}$ is the smallest distance between the boundaries of neighbor cylinders or spheres.

At large $k$, the theoretical probability distributions $\tilde{P}_\mathrm{A}(k)$ and $\tilde{P}_\mathrm{B}(k)$ for the different structure types are described by the asymptotes, which are the first terms ($j = 1$) in the expansions (19), (27), (30), (34), and (37). The slope of the linear dependence of $\ln \tilde{P}_\alpha(k)$ on $k$ (the inset in figure 5) is inversely proportional to the square of the characteristic spatial size of the domain. The external domains of cylinders and balls (figure 1b,c) are approximated in the theoretical model by the symmetrical domains in the shape of cylindrical and spherical layers, and the width of those layers can be taken as the characteristic domain size.

In the computer simulations, the probability distributions $\tilde{P}_B(k)$ for the external domains are found directly. To estimate the effective outer radius $r_{\text{out}}^{(\text{eff})}$ for such domains, the slope of the dependences of $\ln \tilde{P}_\alpha(k)$ on $k$ is fitted by the corresponding theoretical solutions. For the structure of the cylinders of radius $R = 4a$ with the period $d = 15a$, the best fit is provided by the value $r_{\text{out}} = 13.1a$ (figure 5), which lies between $d - R$ and $d$. For the structure of balls, three dependences are fitted for the different values of $d$ (figure 10), all of them are within the range from $d_s - R_s$ to $d_s$, and we see that $r_{\text{out}}^{(\text{eff})}$ is not directly proportional to $d$.

In the limit of large values of period ($d/R$, $d_s/R_s >> 1$), the effective value of $r_{\text{out}}^{(\text{eff})}$ can be approximated with the power-law dependence. Consider a ball in the center of the first coordination sphere of radius $d_s$. The fraction $\varphi_s$ of the sphere domain covered by other balls is proportional to $\varphi_s \sim R_s^2 / d_s^2$, the number of the coordination spheres in the ball of radius $r_{\text{out}}^{(\text{eff})}$ is equal to $m \approx r_{\text{out}}^{(\text{eff})} / d_s$. Assuming that the fraction $\varphi_s$ is approximately the same for all coordination spheres, the solid angle covered by other balls from the point of view in the center of a chosen ball is equal to $m\varphi_s$. Let $r_{\text{out}}^{(\text{eff})}$ be a distance at which any random trajectory will be finally absorbed, then $m\varphi_s \approx 1$. For the system of cylinders, the first coordination cylinder of radius $d$ should be considered, with the surface fraction $\varphi \sim R^2 / d^2$ covered by other cylinders. Similar reasons lead to the condition $m\varphi \approx 1$, where $m \approx r_{\text{out}}^{(\text{eff})} / d$. Then, the effective outer radius in the limit of large $d$

$$r_{\text{out}}^{(\text{eff})} \sim \frac{d^2}{R} \text{ (cylinders)}, \quad r_{\text{out}}^{(\text{eff})} \sim \frac{d_s^3}{R_s^2} \text{ (balls)}. \tag{39}$$

The dependences of number average block lengths $\bar{k}_{n\alpha}$ and dispersities $Đ_\alpha$ on the domain sizes are also presented for all the structure types (figures 3, 6-8, 11-13). The solid curves describe the results of the analytical calculations and the markers describe the results of the computer simulations. For a lamella (figure 3), a cylinder (figure 6), and a ball (figure 11), the average block length $\bar{k}_{nA}$ and the dispersity $Đ_A$ grow practically linearly with the size of the corresponding domain ($L$, $R$, and $R_s$, respectively) both in the theory and in the computer simulations. The linear law (12) and (13) follows from the Levy-flight statistics $P_\alpha(k) \sim 1/k^{3/2}$ at the moderate values of $k$ (9) together with the exponential decrease (10) beginning from the unique characteristic size $k^* \approx (d_A/a)^2$.

However, the values of $\bar{k}_{nA}$ found in the computer simulations are remarkably larger than those calculated numerically (see, for example, figure 6 for cylinders). This difference can be explained by the peculiarities of the analytical solution for the probability distribution $P_A(k)$ at small values of block length $k$. Chain trajectories are described by the Green function (Eq. (1)) as continuous ones, whereas a set of monomer unit coordinates is discrete. It can be expected that the continuous description is not adequate at small $k$. Indeed, at $k = 1$ the probability $\tilde{P}_A(1) \approx 0.42$ in theory and 0.32 in the computer simulations for the ball of radius $R = 4a$. Since the fraction of short blocks of size $k = 1$ and 2 is overestimated in the analytical solution, the fraction of other blocks is underestimated thus leading to smaller values of the number average block size $\bar{k}_{nA}$ in the analytical solution than in the computer simulations.

For the values of the mass average block size $\bar{k}_{w\alpha}$ and dispersity $Đ_\alpha$, the largest contribution is made by the longest blocks. In the computer simulations, a block trajectory can be interrupted by a chain end before it crosses the interface. A lack of long blocks in the computer simulations in comparison with the analytical solution leads to somewhat smaller values of the dispersity (figures 3, 6, and 11).

For B blocks, the dependences of number average block length $\bar{k}_{nB}$ on the outer radius $r_{out}$ and inner radius ($R$ or $R_s$) obtained in the analytical calculations are much smoother than the dependences of $\bar{k}_{nB}$ on the size of the external domain (spatial period or radius) obtained in the computer simulations (figures 12a, 13a, 15a, and 16a). In the analytical calculations, the dependences are practically linear in the presented cases. In the computer simulations, $\bar{k}_{nB}$ grows faster than linearly with the structure period $d$ (or $d_s$) under the fixed cylinder (or ball) radius. For a block beginning near a given cylinder, the nearest points of the surfaces of the other cylinders move away with increasing $d$, and simultaneously other cylinders become rarefied in space. In the result, the effective outer radius should change faster than linearly with $d$. And the same conclusion is valid for the structure of balls arranged in the lattice.

The dispersity $Đ_B$ changes practically linear with the period $d$ ($d_s$) and with the radius $R$ ($R_s$) at moderate values of the thickness $d - R$ in the computer simulations. For larger values, the dispersity sharply decreases mainly because of a decrease in the values of mass average block length $\bar{k}_{wB}$, which are sensitive to the lack of long blocks. For example, there is a certain fraction of the homopolymer B chains of length 1000 in the simulation box for the balls of radius $R_s = 3a$ with the structure period $d_s = 20a$. Therefore, in the limit $N \to \infty$ a noticeable fraction of longer blocks is expected to be present in this system. The value of $Đ_B$ calculated from the expressions

(24) and (33) changes approximately linearly with $r_{out}$ and $R$ (thick curves in figures 7b and 8b) at the comparable values of the cylinder radius $R$ and the thickness $r_{out} - R$. For the length distribution (40) of blocks in the spherical layer with the inner radius $R_s$ and outer radius $r_{out}$, the dependence is also close to linear (figure 12b and 13b). The dispersity is quite large in all considered cases and, besides, it is predicted to grow with the domain size (estimation (13)). That is, the dispersities can considerably exceed those for the polymers synthesized via radical polymerization, which is characterized by the most probable (Flory) distribution with the dispersity equal to 2.

Despite a very broad distribution of block lengths, the polymers obtained via the conformation-dependent sequence design are able to form stable segregated structures. The stability of globules with a hydrophobic core and a polar shell formed by protein-like copolymers was demonstrated in the computer simulations and in the theory [7, 8] and also experimentally [9, 10]. The formation of lamellar structure in the melt of multiblock copolymers prepared via the design with a lamellar pattern was also observed in the computer simulations [15]. After switching on the repulsive interactions between monomer units of different types, the lamellae with the period considerably larger than that of the initial pattern were formed. It can be expected that other patterns can also be reproduced for the other types of the designed block copolymers in melt.

It is worth to note that very short blocks can be located in the domains of “alien” type or at the interface and, therefore, they hinder the microphase separation in a melt. The effect of short blocks on the microphase separation was studied theoretically for the melt of the random multiblock copolymer of special type [36]. Thus, for better reproduction of a pattern used in the conformation-dependent design, it is desirable to prevent the formation of very short blocks during the modification.

## 4. Conclusion

In the present paper, the statistical characteristics of random multiblock copolymers of the special type have been studied. The type A or B is assigned to the monomer units in accordance with their spatial positions in the homopolymer melt, where the periodic pattern is introduced. The monomer unit sequences designed in such a way depend on the pattern type and the domain sizes (figure 1). The designed random multiblock copolymers are able to reproduce, for example, the “parent” lamellar structure in the melt with repulsive interactions of A and B monomer units, as was demonstrated previously in the computer simulations [15]. We have been considered together the several structure types (alternating lamellae, hexagonally arranged cylinders, and balls packed in the body-centered cubic lattice), which are typical of microphase separation in block copolymer melts.

Such bulk modification of a polymer melt still has not been performed experimentally. As a way to realize the process, we can suggest to initiate a photochemical reaction in a sample creating a light 3D-pattern by optical methods.

For the modified macromolecules, the block length distributions and the average block lengths have been found both in the computer simulations of the melt equilibrated via the DPD method and theoretically using the description of random walk statistics of polymer chains with the diffusion-type equations. When moving along a polymer chain, the monomer unit type changes at the interfaces. Simultaneously, the polymer block is terminated, as described by the absorbing boundaries or Dirichlet-type boundary conditions. For the external domain of balls and of cylinders, the approximation of a spherical layer and a cylindrical one, respectively, have been used in the theory (figure 4).

At moderate length $k$ of blocks, the Levy flight-type distribution of block length $P_\alpha(k) \sim 1/k^{3/2}$ is observed for all the pattern types. This dependence describes the asymptote of the block length distribution for the blocks starting near an absorbing plane in a semi-space. Therefore, this asymptote should be valid for any pattern in 3D until the initial surface curvature reveals itself and the other boundaries become attainable, that is, before the characteristic spatial size $\sqrt{k}a$ of a block reaches the smallest characteristic size of the domain.

The general shape of the block length distributions obtained in the computer simulations and in the theoretical calculations is very similar for the blocks in lamellae, cylinders, and balls. For the external domain of balls and of cylinders, those distributions are not so similar. However, the slope of the dependence found in the computer simulations can be well fitted at large block lengths by a proper choice of the thickness of the spherical or cylindrical layer in the theoretical calculations. The asymptote $P_\alpha(k) \sim \exp\left(-\lambda \frac{ka^2}{d_{\mathrm{as}}^2}\right)$, $\lambda = \mathrm{const}$, can be used at large $k$, where the asymptotic length $d_{\mathrm{as}}$ nonlinearly depends on two parameters, $R$ and $d$ (or $R_\mathrm{s}$ and $d_\mathrm{s}$), for the external domains of cylinders and balls. It can be expected that for more complex lattices such asymptotic length can depend on several geometric parameters that determine the size and arrangement of the domains. For the symmetric domains of cylindrical and spherical layer, the layer thickness is the only relevant parameter.

The average block length and the dispersity depend linearly on the characteristic domain size if this size is unique. In a multi-scaled domain, such dependence can be nonlinear. The numerical values of the average block lengths calculated from the analytical solutions of the diffusion-type equations are somewhat underestimated in comparison with the results of the computer simulations because of the excess number of very short blocks, for which the continuous theoretical description is not very adequate. On the other side, the block length distribution in finite polymer chains in the computer simulations does not describe the statistics of infinite random walks for patterns with a large spatial period. In that case, the effect of chain ends, which terminate the blocks, becomes important for the chains of spatial size comparable with the structure period. Therefore, both methods complement each other and can provide rich information about the statistics of random walks in a patterned media.

### Acknowledgment

The research was financially supported by the Russian Foundation for Basic Research (project no. 19-03-00988).

## References

[1] Runnels C M, Lanier K A, Williams J K, Bowman J C, Petrov A S, Hud N V, and Williams L D 2018 Folding, Assembly, and Persistence: The Essential Nature and Origins of Biopolymers. *J. Mol. Evol.* **86** 598-610

[2] Barrett C, Huang F W, Reidys C M 2017 Sequence–structure relations of biopolymers. *Bioinformatics* **33** 382–389

[3] *Introduction to Protein Structure Prediction: Methods and Algorithms (ed. by Rangwala H and Karypis G)* 2010 (Hoboken, NJ: John Wiley & Sons)

[4] Pande V S, Grosberg A Y, Tanaka T 2000 Heteropolymer freezing and design: Towards physical models of protein folding. *Reviews of Modern Physics* **72** 259-314

[5] Khokhlov A R, Khalatur P G 1999 Conformation-Dependent Sequence Design (Engineering) of AB Copolymers. *Phys. Rev. Lett.* **82**, 3456

[6] Govorun E N, Ivanov V A, Khokhlov A R, Khalatur P G, Borovinsky A L, Grosberg A Yu 2001 Primary sequences of proteinlike copolymers: Levy-flight-type long-range correlations. *Physical Review E* **64** R40903

[7] Govorun E N, Khokhlov A R, and Semenov A N 2003 Stability of dense hydrophobic-polar copolymer globules: Regular, random and designed sequences. *Eur. Phys. J. E* **12** 255-264

[8] van den Oever J M P, Leermakers F A M, Fleer G J, Ivanov V A, Shusharina N P, Khokhlov A R, and Khalatur P G 2002 Coil-globule transition for regular, random, and specially designed copolymers: Monte Carlo simulation and self-consistent field theory. *Phys. Rev. E* **65** 041708

[9] Lozinsky V I 2006 The approaches to chemical synthesis of protein-like copolymers. *Adv. Polym. Sci.* **196** 87-127

[10] Lozinsky V I, Simenel I A, Kulakova V K, Kurskaya E A, Babushkina T A, Klimova T P, Burova T V, Dubovik A S, Grinberg V Y, Galaev I Y, Mattiasson B, Khokhlov A R 2003 Synthesis

and studies of N-vinylcaprolactam/N-vinylimidazole copolymers that exhibit the “proteinlike” behavior in aqueous media. *Macromolecules* **36** 7308

[11] Chertovich A V, Govorun E N, Ivanov V A, Khalatur P G, and Khokhlov A R 2004 Conformation-dependent sequence design: evolutionary approach. *Eur. Phys. J. E* **13** 15-25

[12] Denesyuk N A and Erukhimovich I Ya 2000 Adsorption of a correlated random copolymer chain at a liquid-liquid interface. *J. Chem. Phys.* **113** 3894

[13] Zheligovskaya E A, Khalatur P G, Khokhlov A R 1999 Properties of AB copolymers with a special adsorption-tuned primary structure. *Phys. Rev. E* **59** 3071

[14] Jhon Y K, Semler J J, Genzer J, Beevers M, Guskova O A, Khalatur P G, Khokhlov A R 2009 Effect of Comonomer Sequence Distribution on the Adsorption of Random Copolymers onto Impenetrable Flat Surfaces. *Macromolecules* 42 2843-2853

[15] Govorun E N, Gavrilov A A, Chertovich A V 2015 Multiblock copolymers prepared by patterned modification: Analytical theory and computer simulations. *J. Chem. Phys.* **142** 204903

[16] MacEwan S R, Weitzhandler I, Hoffmann I, Genzer J, Gradzielski M, and Chilkoti A 2017 Phase behavior and self-assembly of perfectly sequence-defined and monodisperse multi-block copolypeptides. *Biomacromolecules* **18** 599–609

[17] Behl M, Balk M, Lützow K, Lendlein A 2021 Impact of block sequence on the phase morphology of multiblock copolymers obtained by high-throughput robotic synthesis. *Eur. Polym. J.* **143** 110207

[18] Gringolts M L, Denisova Yu I, Finkelshtein E Sh, Kudryavtsev Y V 2019 Olefin metathesis in multiblock copolymer synthesis. *Beilstein J. Org. Chem.* **15** 218-235

[19] Antonopoulou M N, Whitfield R, Truong N, Anastasaki A 2020 Concurrent control over sequence and dispersity in multiblock copolymers. *ChemRxiv* doi.org/10.26434/chemrxiv.12895754.v1

[20] Gody G, Maschmeyer T, Zetterlund P B, and Perrier S 2013 Rapid and quantitative one-pot synthesis of sequence-controlled polymers by radical polymerization. *Nature Communications* **4** 2505

[21] Grosberg A Y and Khokhlov A R 1994 *Statistical Physics of Macromolecules* (Woodbury, NY: AIP)

[22] Feller W 1971 *An Introduction to Probability Theory and Its Applications*, vol. II (3rd ed.) (New York: John Wiley & Sons)

[23] Redner S 2001 *A Guide to First-Passage Processes* (Cambridge University Press)

[24] Bray A J, Majumdar S N, and Schehr G 2013 Persistence and First-Passage Properties in Non-equilibrium Systems. *Advances in Physics* 62 225-361

[25] Ben-Naim E and Krapivsky P L 2010 First-passage exponents of multiple random walks. *J. Phys. A: Math. Theor.* **43** 495008

[26] Avram F and Perez-Garmendia J-L 2019 A Review of First-Passage Theory for the Segerdahl-Tichy Risk Process and Open Problems. *Risks* **7** 117

[27] Simon D and Derrida B 2008 Quasi-Stationary Regime of a Branching Random Walk in Presence of an Absorbing Wall. *Journal of Statistical Physics* **131** 203-233

[28] Jung Y, Kim Y 2002 Scaling and Survival Properties of Random Walks with Absorbing and Moving Walls. *J Korean Phys. Soc.* **141** 17-170

[29] Halpern L, Rauch J 1995 Absorbing boundary conditions for diffusion equations. *Numer. Math.* **71** 185–224

[30] Mandelbrot B B 1982 *The Fractal Geometry of Nature* (Updated and augm. ed.) (New York: W. H. Freeman)

[31] Budak B M, Samarskii A A, and A N Tikhonov A N 1964 *A collection of problems on mathematical physics* (Pergamon Press)

[32] Hoogerbrugge P J and Koelman J M V A 1992 *Europhys. Lett.* **19** 155.

[33] Groot R D and Warren P B 1997 *J. Chem. Phys.* **107** 4423

[34] Allen M P and Tildesley D J 1987 *Computer Simulation of Liquids* (Oxford: Clarendon Press).

[35] Watson G N 1944 *A treatise on the theory of Bessel functions* (Cambridge University Press)

[36] Govorun E N and Chertovich A V 2017 Microphase Separation in Random Multiblock Copolymers *J. Chem. Phys.* **146** 034903